\documentclass[fleqn,usenatbib]{mnras}

\usepackage{newtxtext,newtxmath}
\usepackage{comment}
\usepackage[T1]{fontenc}

\DeclareRobustCommand{\VAN}[3]{#2}
\let\VANthebibliography\thebibliography
\def\thebibliography{\DeclareRobustCommand{\VAN}[3]{##3}\VANthebibliography}

\usepackage{graphicx}	
\usepackage{amsmath}	
\usepackage{xcolor} 
\usepackage{caption}
\usepackage{subcaption}
\usepackage{lineno}

\title[The kiloparsec-scale jet of 3C\,306]{The evolution of 3C\,306 in the galaxy cluster RXCJ1454.4+1622: linking spectral ageing, ICM conditions, and jet dynamics through MHD simulations}

\author[A. S. R. Antas et al.]{
A. S. R. Antas,$^{1,2}$\thanks{E-mail: abraao.antas@uansp.edu.br (ASRA)}
A. Caproni,$^{1}$,
L. Loinard,$^{3,4}$
G. S. Souza$^{1}$,  
T. F. Laganá$^{1}$
\\
$^{1}$N\'ucleo de Astrof\'\i sica, Universidade Cidade de S\~ao Paulo, R. Galv\~ao Bueno 868, Liberdade, S\~ao Paulo, SP, 01506-000, Brazil\\
$^{2}$Centro Universitário Adventista de São Paulo, Rua Pr. Hugo Gegembauer, 265, 13184-010, Hortolândia, SP, Brazil\\
$^{3}$Instituto de Radioastronomía y Astrofísica, Universidad Nacional Autónoma de México, Morelia 58341, México\\
$^{4}$1Black Hole Initiative at Harvard University, 20 Garden Street, Cambridge, MA 02138, USA
}

\date{Accepted XXX. Received YYY; in original form ZZZ}

\pubyear{2024}

\begin{document}
\label{firstpage}
\pagerange{\pageref{firstpage}--\pageref{lastpage}}
\maketitle

\begin{abstract}
We present a methodology for analysing radio sources in galaxy clusters by combining multi-frequency radio observations, X-ray analysis of the intracluster medium (ICM), and three-dimensional magnetohydrodynamic (MHD) simulations, and apply it to the galaxy cluster RXCJ1454.4+1622 and its associated FR\,II radio galaxy 3C\,306 (IC\,4516). The radio emission exhibits a curved structure connecting the lobes, likely indicating a complex dynamical evolution influenced by environmental effects. Spectral index and break-frequency maps reveal a relatively homogeneous radiative ageing distribution, with characteristic spectral ages of $\sim 220$~Myr and a smooth spatial gradient. From equipartition arguments, we estimate an equivalent jet magnetic field of $2.1 \pm 0.3~\mu$G, consistent with values observed in FR\,II radio galaxies. These observational constraints are used to initialise non-relativistic MHD simulations exploring three scenarios: jet precession, bulk motion of the source through the ICM, and a combined model including both effects. Neither pure precession nor pure source motion alone fully reproduces the observed morphology, whereas the combined model provides a more consistent representation of the jet curvature and lobe asymmetries. Nevertheless, the simulations indicate that source motion is the dominant mechanism shaping the large-scale radio morphology, with jet precession playing a secondary role. The inferred source velocity ($\sim 280$~km\,s$^{-1}$) and precession timescale ($\sim 250$~Myr) are compatible with the spectral ageing constraints. These results suggest that the morphology of 3C\,306 is influenced by the interplay between jet precession and dynamical interaction with the ICM, highlighting the role of environmental effects in radio-galaxy evolution in clusters.

\end{abstract}




\section{Introduction}

Active Galactic Nuclei (AGN) are among the most energetic phenomena in the Universe, powered by the accretion of matter onto supermassive black holes (SMBHs) located at the centres of galaxies. Their emission extends across the entire electromagnetic spectrum, from radio waves to gamma rays, and, in some cases, includes relativistic jets that can extend far beyond the host galaxy, sometimes reaching megaparsec scales (e.g., \citealt{2019ARA&A..57..467B, 2023JApA...44...13D}). Understanding these jets is essential for elucidating not only the physical processes near SMBHs — including the mechanisms of jet launching and collimation — but also their impact on the surrounding interstellar and intergalactic media.

The interaction between relativistic jets and the interstellar medium (ISM) can influence star formation rates \citep[e.g.,][]{2012ApJ...757..136W,2011Sci...333..182H}, redistribute gas within galaxies \citep[e.g.,][]{2011ApJ...728...29W}, and regulate the growth of the central SMBH through feedback processes \citep[e.g.,][]{2006MNRAS.365...11C,2011ApJ...737...23Y}. On larger scales, AGN jets deposit significant amounts of energy into the surrounding intergalactic and intracluster medium (IGM/ICM), inflating cavities, driving shocks, and heating the ambient gas. These processes play a key role in regulating cooling flows and in shaping the thermodynamical evolution of galaxy clusters \citep[e.g.,][]{2005Natur.433...45M,2007ARA&A..45..117M,2012NJPh...14e5023M}.

The radio galaxy 3C\,306 (IC\,4516) is hosted in the galaxy cluster RXCJ1454.4+1622. It exhibits a classical edge-brightened, double-lobed structure characteristic of FR\,II radio galaxies, as reported in radio surveys and catalogues \citep[e.g.][]{1974MNRAS.167P..31F,2017A&A...601A..81C}. Beyond this general classification, however, its radio emission shows clear deviations from axial symmetry, including curved jets and distorted lobes, indicating that environmental effects play an important role in shaping its large-scale structure.

Such distortions are commonly interpreted as the result of interactions between relativistic jets and the surrounding intracluster medium (ICM). In particular, mechanisms such as jet precession \citep[e.g.,][]{1982ApJ...262..478G,2004ApJ...602..625C,2013MNRAS.428..280C,2017ApJ...851L..39C,2019MNRAS.482..240K,2021ApJ...908..178N}, bulk motions of the host galaxy within the cluster potential \citep[e.g.,][]{1972ApJ...176....1G,1985AJ.....90..954O,1987ApJ...316..113O,1993ApJ...408..428O,2024MNRAS.533.1341A}, or large-scale ICM flows induced by cluster mergers \citep[e.g.,][]{2000MNRAS.311..649S,2002NewAR..46..135B} can give rise to distorted radio structures, including bent and S-shaped morphologies. The observed structure of 3C\,306 is consistent with the action of more than one of these processes, making it an interesting system for investigating the dynamical interplay between AGN jets and the ICM.

A widely used diagnostic of the evolution of radio sources is provided by spectral ageing analysis. The synchrotron emission from relativistic electrons progressively steepens due to radiative losses, primarily synchrotron emission and inverse-Compton scattering off cosmic microwave background photons \citep{1962SvA.....6..317K,1973A&A....26..423J}. By modelling the spatial variations of the radio spectrum, it is possible to estimate the spectral age of the electron population, which provides a lower limit to the time elapsed since the last significant particle acceleration event \citep[e.g.,][]{1999A&A...345..769M,2013MNRAS.435.3353H}. In this context, spatially resolved spectral ageing maps provide useful constraints on the dynamical and radiative evolution of radio jets.

In parallel, numerical simulations have become an important tool for interpreting the complex structures observed in radio galaxies. Hydrodynamic (HD) and magnetohydrodynamic (MHD) simulations have shown that jet propagation, stability, and large-scale morphology are strongly influenced by interactions with the ambient medium, including density gradients, external flows, and jet precession \citep[e.g.,][]{1991MNRAS.250..581F,2007MNRAS.380...51K,2024MNRAS.533.1341A}. In particular, MHD simulations are well suited to capture the role of magnetic fields in jet collimation, shock formation, and particle acceleration processes. By comparing synthetic morphologies derived from simulations with observational data, it is possible to place constraints on the physical mechanisms responsible for the observed structures.

In this work, we combine X-ray analysis of the ICM, multi-frequency radio observations,  and non-relativistic MHD simulations to investigate the processes shaping the distorted morphology of 3C\,306. In particular, we explore whether jet precession, bulk motion of the source, or a combination of both can reproduce the observed radio structure, and how these mechanisms relate to the physical conditions of the surrounding ICM.

This paper is structured as follows. In \autoref{sec:xraydata}, we present the X-ray analysis of the galaxy cluster RXCJ1454.4+1622, including the procedures adopted to derive its temperature and density profiles, which provide the basis for constraining the ICM conditions used in the MHD simulations. The radio data analysis of 3C\,306, including the estimation of spectral indices, break frequencies, equivalent magnetic field, and spectral age maps, is presented in \autoref{sec:radiodata}. The numerical setup of the MHD simulations—covering the initial conditions, jet injection parameters, and magnetic field configuration—is described in \autoref{sec:NumSim}. The simulation results for the three scenarios (pure precession, pure galactic motion, and a composite model) are presented in \autoref{sec:SimRes}, where they are compared with the observed radio morphology. The implications of these results for jet–ICM interactions and AGN dynamics are discussed in \autoref{sec:Discussion}. Finally, we summarise our conclusions in \autoref{sec:conclusions}.

Throughout this study, we adopt a $\Lambda$CDM cosmology with $H_0 = 71$~km\,s$^{-1}$\,Mpc$^{-1}$, $\Omega_\mathrm{M} = 0.27$, and $\Omega_\Lambda = 0.73$, and assume a redshift of $z = 0.04528$ \citep{Albareti_2017} for 3C\,306, corresponding to a spatial scale of 0.8785~kpc arcsec$^{-1}$.

\section{The X-Ray data} \label{sec:xraydata}

\subsection{XMM-Newton data reduction}
The data were reduced using the XMM-Newton Science Analysis System (SAS; \citealt{2004ASPC..314..759G}), version 21.0, with calibration files updated to November 2019. To filter periods affected by solar flares, a $2\sigma$ clipping procedure was applied to the light curves in the energy band [2.5–12.0] keV. 

To account for the background contribution from each detector, we extracted a background spectrum from an outer annular region of the observation in the [10.0–12.0] keV energy band. These spectra were then compared with blank-sky background data from \cite{Read2003}, using the same detector region and energy range. The observational background was subsequently normalised to the blank-sky background, and the resulting normalisation parameter was used in the spectral fitting.

\subsection{X-ray spectral analysis} \label{x-raySA}

We extracted cluster spectra in concentric annuli around the X-ray emission peak. To ensure a good signal-to-noise ratio (S/N), we assumed a threshold of 900 counts in each ring, which corresponds to an S/N $\ge 30$ \citep{Lagana_Gsouza2019}. We excluded from our analysis the brightest points that were located within the field of view of the detectors. These points were identified using the \textit{edetect-chain} task that identifies X-ray sources in images taken with the EPIC instruments, which include the MOS and PN cameras.

For spectral fit, we used XSPEC version 12.12.1 \citep{Arnaud1996} and modelled the obtained spectra with a MEKAL single-temperature plasma emission model \citep{Kaastra1993}. The free parameters are X-ray temperature and metallicity of the ICM in RXCJ1454.4+1622. Spectral fit were performed in the energy range of [0.7-7.0] keV, and we also excluded the energy range of 1.2 to 1.9 keV to avoid any influence of the Al and Si instrumental lines. For the Galactic absorption, we adopted the total hydrogen column density ($N_{\rm H}$), computed as the sum of the atomic hydrogen column density from the Leiden/Argentina/Bonn (LAB) survey \footnote{\url{https://cade.irap.omp.eu/dokuwiki/doku.php?id=lab}} \citep{2005A&A...440..775K} and the molecular hydrogen contribution, based on the prescription of \citet{2013MNRAS.431..394W}.

\begin{figure}
	\includegraphics[scale=0.495]{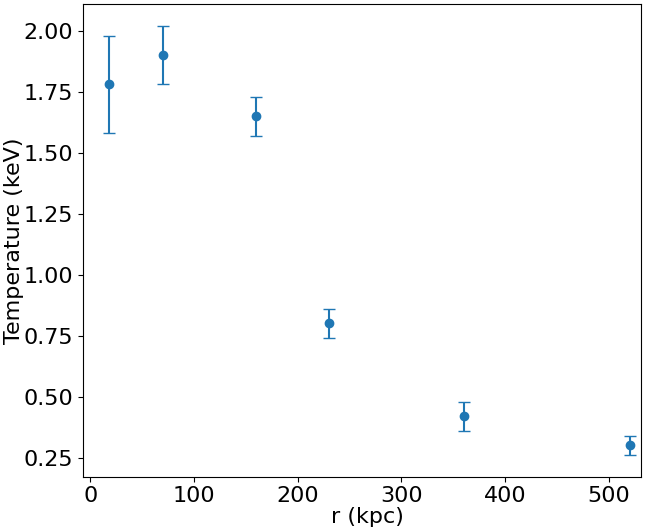}
	\centering
    \caption{Radial profile of the temperature distribution in the galaxy cluster RXCJ1454.4+1622 calculated in this work from XMM-Newton observations.}
    \label{fig:TempPerfil}
\end{figure}

In \autoref{fig:TempPerfil}, we present the radial temperature profile of the galaxy cluster RXCJ1454.4+1622 obtained from the spectral fitting of its XMM-\textit{Newton} data.
To our knowledge, this is the first estimate of such profile for this galaxy cluster made in the literature. The temperature profile shows a peak ($\sim 1.9$ keV) around 80 kpc and a systematic decrease for larger radial distances. Below 80 kpc, the temperature is slightly lower at $\sim 20$ kpc, suggesting that RXCJ1454.4+1622 consistent with a cool-core system.

\subsection{Radial density profile of RXCJ1454.4+1622}

The large-scale structure of the ICM in RXCJ1454.4+1622 was investigated using the radial X-ray surface-brightness distribution derived from XMM-Newton EPIC observations in the $0.5$--$2.0$ keV band. The analysis was based on the EPIC photon image and the corresponding exposure map, from which an exposure-corrected image was constructed by excluding low-exposure regions and dividing the count image by the exposure map. This procedure provides a surface-brightness map suitable for characterising the diffuse cluster emission on large scales.

A key step in this analysis is the definition of the cluster centre. Rather than adopting the brightest pixel of the X-ray map, we determined the centre from the diffuse emission using an iterative, flux-weighted centroiding procedure applied to the exposure-corrected image. In this approach, point sources detected by the XMM-Newton Pipeline Processing System (PPS) were masked, and an additional manual exclusion region was applied to a bright contaminating source south-west of the cluster core. To reduce the impact of the central AGN and other compact structures, the centroid was computed from the diffuse emission within an annulus defined by an inner exclusion radius of 40~kpc and an outer radius of 300~kpc. The centroid was then refined iteratively, converging after five iterations. 

\begin{figure}
    \centering
    \includegraphics[width=\columnwidth]{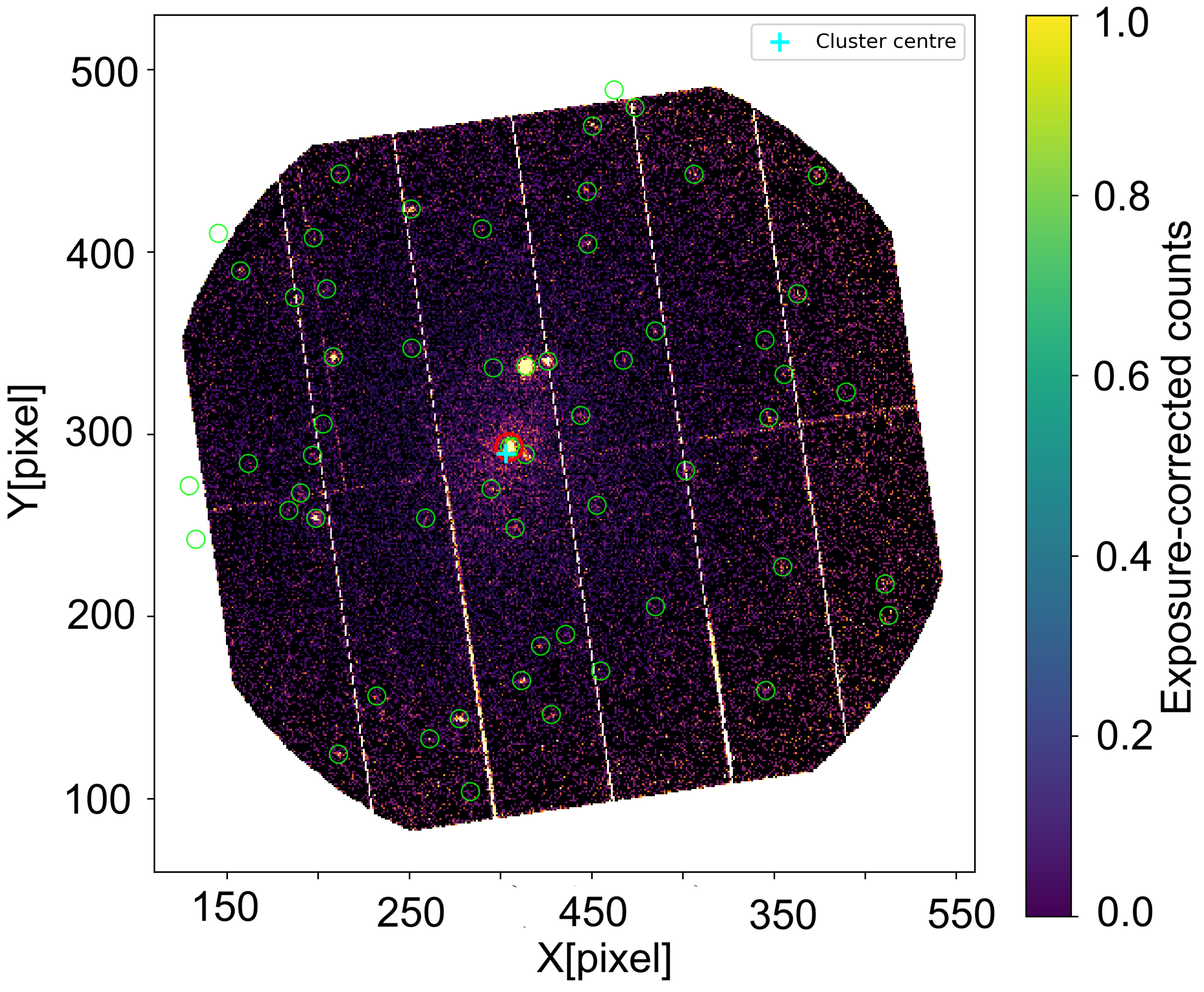}
    \caption{Exposure-corrected EPIC count image of RXCJ1454.4+1622. The cyan cross marks the adopted cluster centre, while the green circles indicate masked compact sources identified from the PPS catalogue, including an additional manual mask (red) applied to a bright contaminating source.}
    \label{fig:xray_image}
\end{figure}

The adopted centre, shown in \autoref{fig:xray_image}, is located at RA = $223.59969 \pm 0.03584$ deg and Dec. = $16.35074 \pm 0.04389$ deg, corresponding to the refined centroid of the extended ICM emission.

\begin{figure}
    \centering
    \includegraphics[width=\columnwidth]{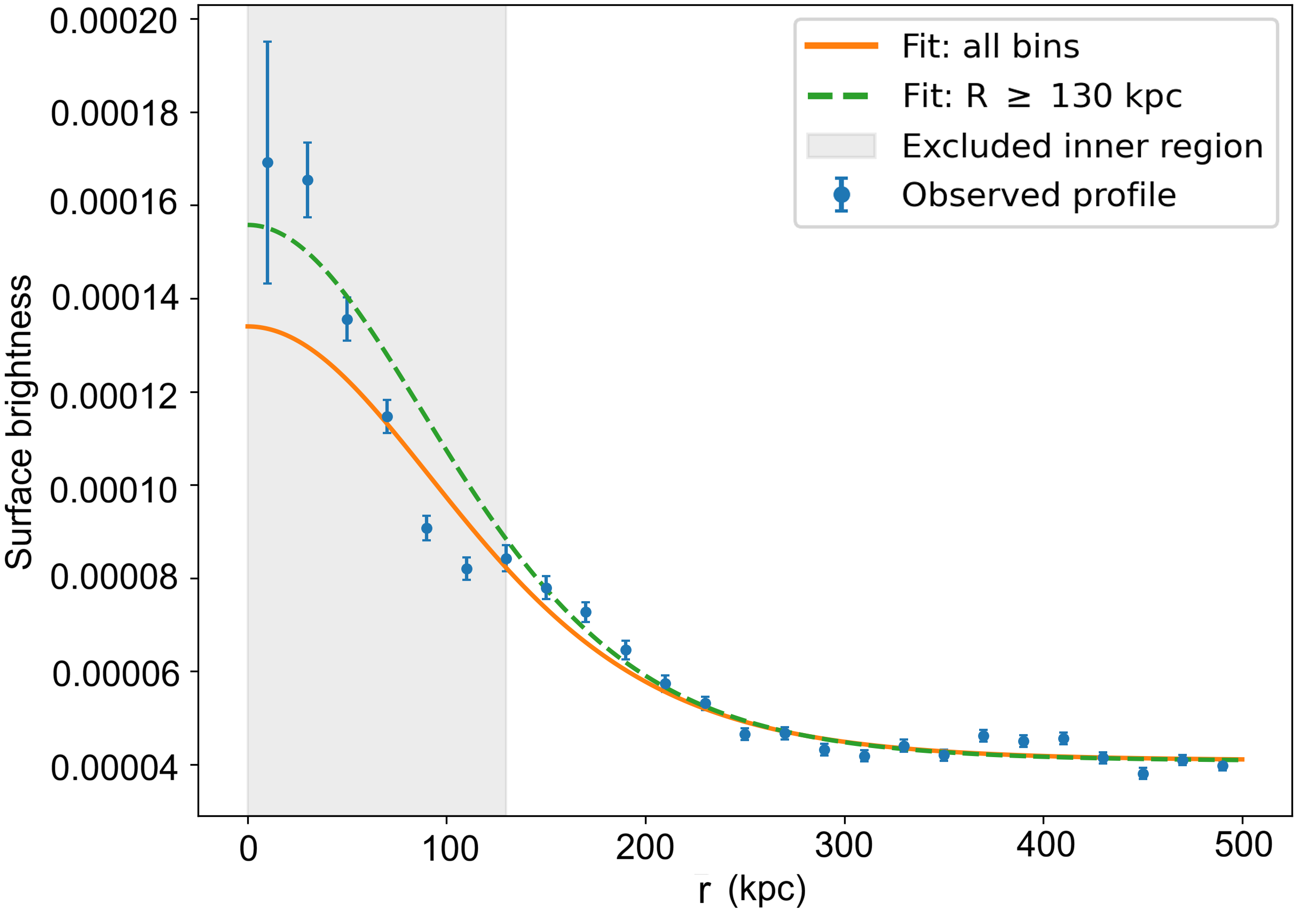}
    \caption{Radial surface-brightness profile of RXCJ1454.4+1622. The shaded region marks the inner $R < 130$ kpc excluded in the restricted fit. The solid and dashed curves correspond to the best-fitting projected $\beta$-model using all radial bins and excluding the central region, respectively.}
    \label{fig:sb_profile}
\end{figure}

With the cluster centre defined, the radial surface-brightness profile was extracted in concentric annuli of width 20~kpc, extending out to 500~kpc. For each annulus, we computed the mean exposure-corrected surface brightness from the unmasked pixels. The associated uncertainty was estimated from the dispersion of pixel values within each annulus, providing an empirical estimate of the radial variability. The resulting profile is shown in \autoref{fig:sb_profile}, where the surface brightness declines smoothly with radius outside the central region.

To model the large-scale gas distribution, we fitted the observed profile with the standard projected $\beta$-model \citep{2009A&A...500..103A},

\begin{equation}
S(R) = S_0 \left(1 + \frac{R^2}{r_c^2} \right)^{-3\beta + 1/2} + B,
\end{equation}
\\where $S_0$ is the central surface-brightness normalisation, $r_c$ is the core radius, $\beta$ is the slope parameter, and $B$ is a constant background term.  

In order to reduce parameter degeneracy and maintain consistency with previous structural analyses, we fixed the slope parameter to $\beta = 1.47$, following \citet{2022MNRAS.513.3013Y}, while $r_c$, $S_0$, and $B$ were left free. We note that this choice reflects a representative value derived from morphological studies and is intended to provide a first-order description of the large-scale gas distribution rather than a fully unconstrained fit.

Because the innermost cluster region shows deviations from the smooth large-scale profile, we explored two fitting strategies: one using all radial bins and another excluding the central region within $R < 130$~kpc. The second approach was adopted to reduce the effect of central excess emission and residual contamination associated with the core. As shown in \autoref{fig:sb_profile}, the restricted fit provides a better description of the outer cluster emission, reducing the reduced chi-square from $\chi^2_\nu \approx 5.85$ to $\chi^2_\nu \approx 3.22$. This behaviour indicates that the projected $\beta$-model captures the large-scale ICM structure reasonably well once the central disturbed region is excluded.

The best-fitting parameters for the restricted fit are $r_c = (2.73 \pm 0.10) \times 10^2$ kpc, $S_0 = (1.15 \pm 0.13) \times 10^{-4}$~counts~s$^{-1}$~pixel$^{-1}$, and $B = (4.07 \pm 0.05) \times 10^{-5}$~counts~s$^{-1}$~pixel$^{-1}$, showing that the large-scale X-ray emission is consistent with a relatively broad gas distribution. The large value of the core radius is consistent with a relatively extended gas distribution, which may reflect either a dynamically disturbed state or the influence of additional central processes not described by a single smooth $\beta$-model.

The surface-brightness profile shown in \autoref{fig:sb_profile} can be related to the underlying gas density distribution under the assumption of a $\beta$-model. In this framework, the three-dimensional gas density profile is given by

\begin{equation}
\rho(r) = \rho_0 \left(1 + \frac{r^2}{r_c^2} \right)^{-3\beta/2},
\label{eq:rhobetamodel}
\end{equation}
\\where $\rho_0$ is the central gas mass density. 

Since the imaging analysis constrains only the shape of the profile, the absolute normalisation $\rho_0$ cannot be directly obtained from the surface-brightness fit and would require spectroscopic measurements. Therefore, to illustrate the implied three-dimensional behaviour of the gas distribution, we adopt a representative central particle density of $n_0 \sim 10^{-2}$ cm$^{-3}$ \citep[e.g.,][]{1986RvMP...58....1S,2002ARA&A..40..539R}. Assuming a fully ionised plasma with mean molecular weight $\mu \approx 0.6$, this corresponds to a central mass density of $\rho_0 = \mu m_p n_0 \sim 10^{-26}$ g cm$^{-3}$.

\begin{figure}
    \centering
    \includegraphics[width=\columnwidth]{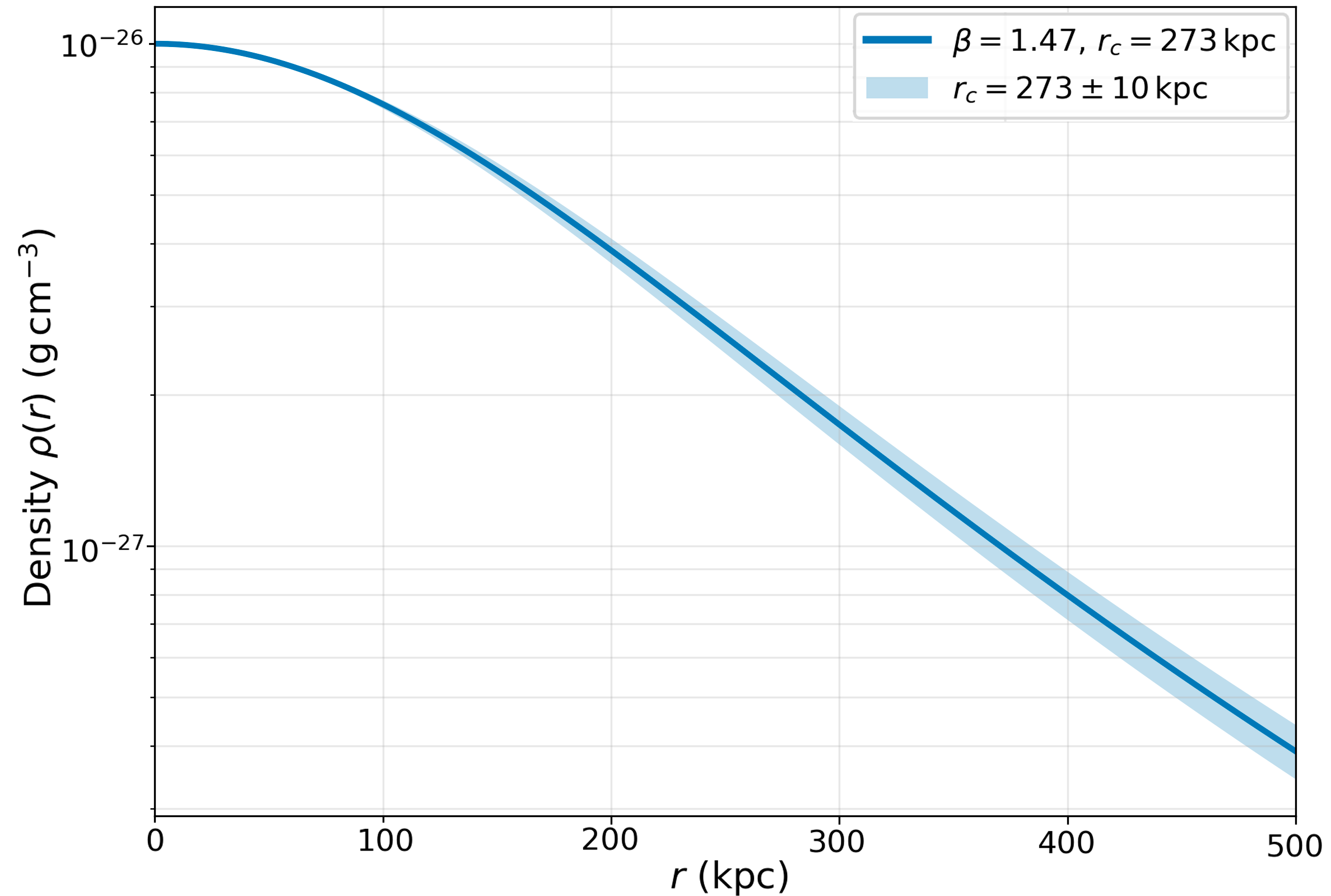}
    \caption{Three-dimensional gas density profile of RXCJ1454.4+1622 derived from the best-fitting projected $\beta$-model. The profile is shown using a representative central density normalisation typical of galaxy clusters.}
    \label{fig:DensityPerfil}
\end{figure}

The resulting three-dimensional density profile is shown in Fig.~\ref{fig:DensityPerfil}. It reproduces the expected radial decline of the ICM density, from values of order $\sim 10^{-26}$ g cm$^{-3}$ in the central region to $\sim 10^{-29}$ g cm$^{-3}$ at $r \sim 500$ kpc.

Thus, the density profile implied by \autoref{eq:rhobetamodel} indicates that the outer ICM of RXCJ1454.4+1622 is reasonably well described by a single smooth $\beta$-model, whereas the central region departs from this behaviour. This central excess might be associated with additional physical processes, such as unresolved AGN-related emission, the presence of a cool-core component (i.e. a central region of higher density and shorter cooling time), or dynamical disturbance in the cluster core, which are not captured by this simplified parametrization.

\section{Radio data analysis} \label{sec:radiodata}

To analyse the radio structure of 3C\,306 from a magnetohydrodynamic perspective, it is first necessary to estimate the age of the source. A commonly employed approach is the spectral ageing analysis, based on the steepening of the synchrotron continuum emission (e.g. \citealt{1985ApJ...291...52M,1999A&A...345..769M,2002NewAR..46..307M,2010A&A...510A..84M, 2017MNRAS.466.2888H}. Although spectral ageing may not provide a precise estimate of the true source age (e.g. \citealt{RUDNICK200295,2001ASPC..250..363B}), it offers a conservative lower limit (e.g., \citealt{1996ASPC..100..281E}) that is enough for constructing an initial model to be tested and refined via MHD simulations.

The radiative ageing of electrons in the lobes of radio-loud galaxies occurs as follows: as the jet propagates through its environment, a strong shock (often referred to as a reverse shock) develops at its terminal point, which in turn accelerates particles to very high energies (Lorentz factors of $\sim 10^5$ or higher) through magnetic inhomogeneities in front of and behind the shock \citep{1978MNRAS.182..147B}. As a simple model, a single population of electrons accelerated to the same energy will leave the acceleration region as the jet advances, travelling downstream. These electrons, filling the radio lobe, will loose energy through inverse Compton, synchrotron, and adiabatic expansion processes. 
 
Under these assumptions and considering a power-law energy distribution with index $\delta$ for the relativistic particles responsible for the synchrotron emission, the standard radio spectrum has a spectral index $\alpha_\mathrm{inj} = (\delta - 1)/2$ \citep{1962SvA.....6..317K} below a critical frequency $\nu_\mathrm{br}$, the break frequency where radiative losses begin to significantly affect the observed spectrum. For frequencies above $\nu_\mathrm{br}$, the spectral index follows $\alpha_\mathrm{h} = \alpha_\mathrm{inj} + 0.5$. If there is no expansion and the magnetic field is constant, the time elapsed since the formation of the source, $t_\mathrm{spec}$, depends on $\nu_\mathrm{br}$, the magnetic field strength, $B$, and the equivalent magnetic field of the cosmic microwave background, $B_\mathrm{CMB} = 3.25(1+z)^2$ $\mu$G, as \citep{2002NewAR..46..307M}:

\begin{equation}\label{eq-tspec}
t_\mathrm{spec} = 1610 \frac{B^{1/2}}{\left(B^2 + B_\mathrm{CMB}^2\right)}\left[(1+z)\nu_\mathrm{br}\right]^{-1/2}\,\textrm{Myr},
\end{equation}
\\where $\nu_\mathrm{br}$ is in units of GHz and $B$ in $\mu$G.

\subsection{Spatially resolved spectral analysis and break frequency estimation}\label{section:ssa}

We constructed a spectral index map of 3C\,306 using RACS observations at 0.89\,GHz and VLA data at 1.50\,GHz. The spectral index $\alpha$ is defined as

\begin{equation}
\alpha = -\frac{\ln (I_0/I_1)}{\ln(\nu_0/\nu_1)},
\end{equation}
\\where $I_0$ and $I_1$ are the intensities at frequencies $\nu_0$ and $\nu_1$, respectively ($\nu_0<\nu_1$). Before performing the spectral analysis, the images were convolved to a common angular resolution and re-gridded to a common pixel scale using the Common Astronomy Software Applications package (CASA; \citealt{2022PASP..134k4501C}). The interpolation was performed using bilinear methods.

\begin{figure*}
\centering
\includegraphics[width=1\textwidth]{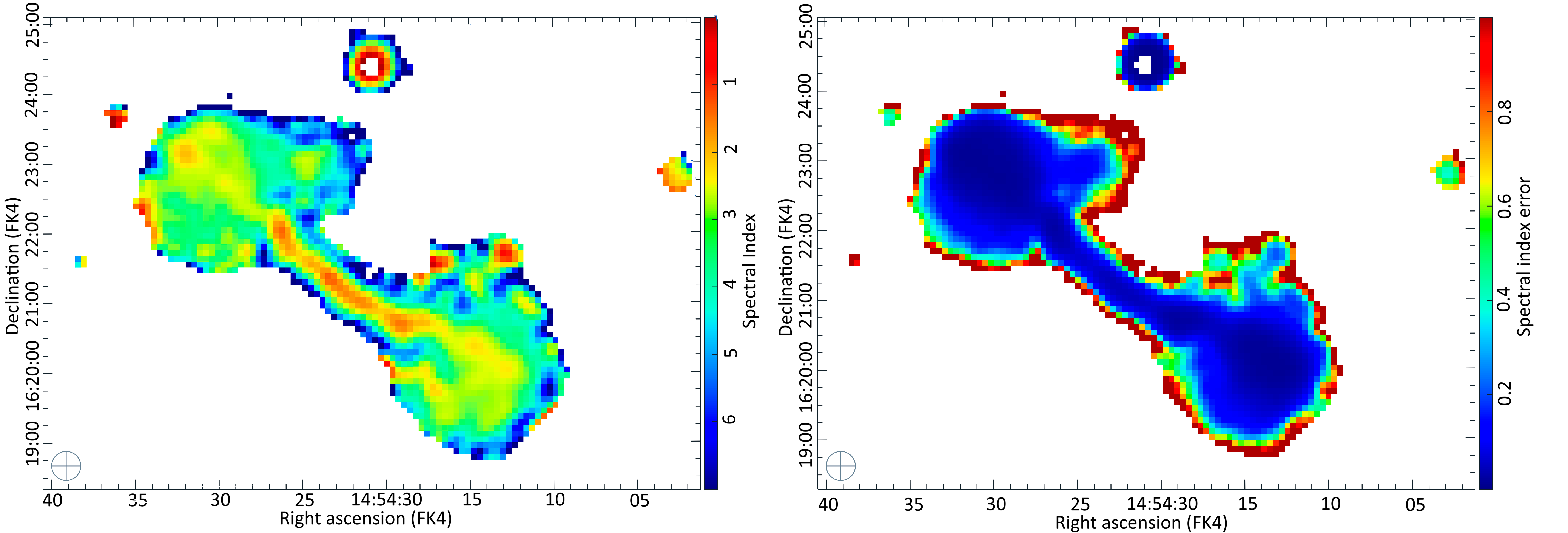}
\caption{Spectral index map derived between 0.89\,GHz (RACS) and 1.50\,GHz (VLA) (left) and corresponding uncertainty map (right) of the radio emission from 3C\,306.}
\label{fig:spix2}
\end{figure*}

\autoref{fig:spix2}  shows the resulting spectral index map (left panel) and the corresponding error map (right panel). The distribution of $\alpha$ reveals two distinct components. The central regions of the lobes exhibit steep spectra, with typical values around $\alpha \sim -3$, consistent with radiative ageing of relativistic electrons in these regions. In contrast, the "bridge" connecting the lobes to the central core displays a flatter spectrum, with $\alpha \sim -0.8$. This flatter region likely traces regions of comparatively less-aged plasma or ongoing particle acceleration or continuous replenishment of relativistic plasma along the jet, while the steeper regions in the lobes reflect accumulated synchrotron and inverse-Compton losses.

The spectral index map reveals a curved jet morphology connecting the two radio lobes. This structure, along with spatial variations in the spectral index along the jet axis --- with flatter spectra (e.g., $\alpha \gtrsim -0.8$) near the core and progressively steeper values farther out --- is consistent with a change in the orientation of the jet axis over time. Similar distortions could arise from a precessing jet scenario in which relativistic plasma is injected along a time-dependent axis. Such directional evolution can be driven by a supermassive binary black hole system \citep[e.g.][]{2000A&A...355..915A,2000A&A...360...57R,Caproni_2006,10.1093/mnras/sts014} or by the Lense--Thirring/Bardeen-Petterson effect \citep{1918PhyZ...19..156L, 1975ApJ...195L..65B}, while relative motion between the jet and the ICM may also induce bending in the jet trajectory \citep[e.g.][]{2024MNRAS.533.1341A}. Alternatively, jet wiggling may be induced by magnetohydrodynamic instabilities, such as Kelvin--Helmholtz \citep[e.g.][]{2006A&A...456..493P}, current-driven kink instabilities \citep[e.g.][]{1998ApJ...493..291B,2000A&A...355..818A,2017MNRAS.469.4957B}, or even by interactions with an inhomogeneous ambient medium \citep[e.g.][]{2008A&A...488..795R,2016MNRAS.461..967M,2017MNRAS.469.4957B}.

The spectral index uncertainty map (Figure~\ref{fig:spix2}, right panel) shows that the regions along the jet axis and in the central parts of the lobes have relatively low uncertainties (typically $\sigma_\alpha \lesssim 0.2$). These areas coincide with high surface brightness regions, which ensures that the derived spectral indices in such locations are reliable.

The derivation of the synchrotron break frequency maps was performed by fitting the radio continuum spectra in each pixel using the full set of publicly available maps, namely TGSS ADR at 0.1475\,GHz \citep{2016yCat..35980078I}, RACS at 0.8875\,GHz and 1.3675\,GHz \citep{2020PASA...37...48M,2023yCatp019003803H}, and VLA at 1.4899\,GHz and 4.8600\,GHz \citep{2011ApJ...739L...1P}. The basis of our analysis relies on the formalism of \citet{2017ApJ...836..174C} and \citet{1999A&A...345..769M}, who developed parametric models to describe compact radio sources with peaked spectra and subsequent high-frequency steepening due to radiative losses. 

The observed flux density $S_\nu$ at frequency $\nu$ is modelled as

\begin{equation}
S_\nu=\frac{S_\mathrm{p}}{1 - e^{-1}}\left(\frac{\nu}{\nu_\mathrm{p}}\right)^{\alpha_{\rm thick}}
\left[1 - e^{-(\nu / \nu_\mathrm{p})^{-(\alpha_\mathrm{thin} + \alpha_{\rm thick})}}\right]
S_{\rm aged}(\nu),
\label{eq:ssa_aged}
\end{equation}
where $S_\mathrm{p}$ is the peak flux density at frequency $\nu_\mathrm{p}$, and $\alpha_{\rm thick}$  and $\alpha_\mathrm{thin}$ are respectively the spectral indices in the optically thick and thin synchrotron regimes.  

The first part of \autoref{eq:ssa_aged}, corresponding to the synchrotron self-absorption (SSA) component, follows the prescription of \citet{2017ApJ...836..174C}. The spectral ageing term $S_{\rm aged}(\nu)$ introduces an additional steepening at $\nu > \nu_\mathrm{br}$ and is parametrized according to \citet{1999A&A...345..769M} as

\begin{equation}
S_{\rm aged}(\nu) = \left(\frac{\nu}{\nu_\mathrm{br}}\right)^{-\alpha_{\rm aged}(\nu/\nu_\mathrm{br})},
\end{equation}
\\with

\begin{equation}
\alpha_{\rm aged}(\nu/\nu_\mathrm{br}) = \frac{1}{2}\left[1 - e^{-\left(\nu/\nu_\mathrm{br}\right)}\right].
\end{equation}

This parametrization provides a smooth transition of the effective spectral slope around $\nu_\mathrm{br}$, as originally proposed by \citet{1999A&A...345..769M}.

For each pixel in the five radio images, we measured the respective flux densities and fitted \autoref{eq:ssa_aged} to them. The fitting procedure employed non-linear least-squares minimization, with physically motivated constraints applied to $S_\mathrm{p}$, $\nu_\mathrm{p}$, and $\nu_\mathrm{br}$. Specifically, $S_\mathrm{p}$ was constrained to lie within the range defined by the minimum and maximum flux densities observed across the input images (scaled by constant factors to avoid numerical instabilities). The characteristic frequencies $\nu_\mathrm{p}$ and $\nu_\mathrm{br}$ were restricted to the range between half of the lowest observing frequency and twice the highest one, ensuring that the fitted values remain within a physically plausible range relative to the data coverage.

The initial value for each parameter was defined as follows: $S_\mathrm{p}$ was taken as the maximum flux across the five bands, $\nu_\mathrm{p}$ as the corresponding frequency of $S_\mathrm{p}$, and $\nu_\mathrm{br}$ was set to the geometric mean of the sampled frequencies. In the fitting procedure, the optically thin and thick spectral indices were fixed to $\alpha_\mathrm{thin} = 0.625$ and $\alpha_{\rm thick} = 2.5$, respectively. The adopted value of $\alpha_\mathrm{thin}$ corresponds to the canonical slope expected from synchrotron-emitting electron populations in the absence of strong spectral curvature \citep[e.g.][]{1962SvA.....6..317K,1979rpa..book.....R}, while $\alpha_{\rm thick}=2.5$ is the asymptotic value predicted for a homogeneous SSA source \citep[e.g.][]{1979rpa..book.....R}. 

To assess the robustness of the derived break-frequency maps with respect to the assumed optically thin spectral index, we repeated the fitting procedure adopting $\alpha_\mathrm{thin} = 0.5$ and $\alpha_\mathrm{thin} = 0.8$, which span the typical range observed in radio-loud AGN. The resulting $\nu_\mathrm{br}$ maps show no significant morphological changes, with the main spatial trends preserved across all tested values. Quantitatively, the relative variations in $\nu_\mathrm{br}$ remain within $\sim$10–20\% over most of the source, with larger deviations confined to low signal-to-noise regions and the outer edges of the emission. This behaviour indicates that the derived break-frequency distribution is robust against reasonable variations of $\alpha_\mathrm{thin}$, supporting the adoption of $\alpha_\mathrm{thin} = 0.625$.

\begin{figure*}
\centering
\includegraphics[width=1\textwidth]{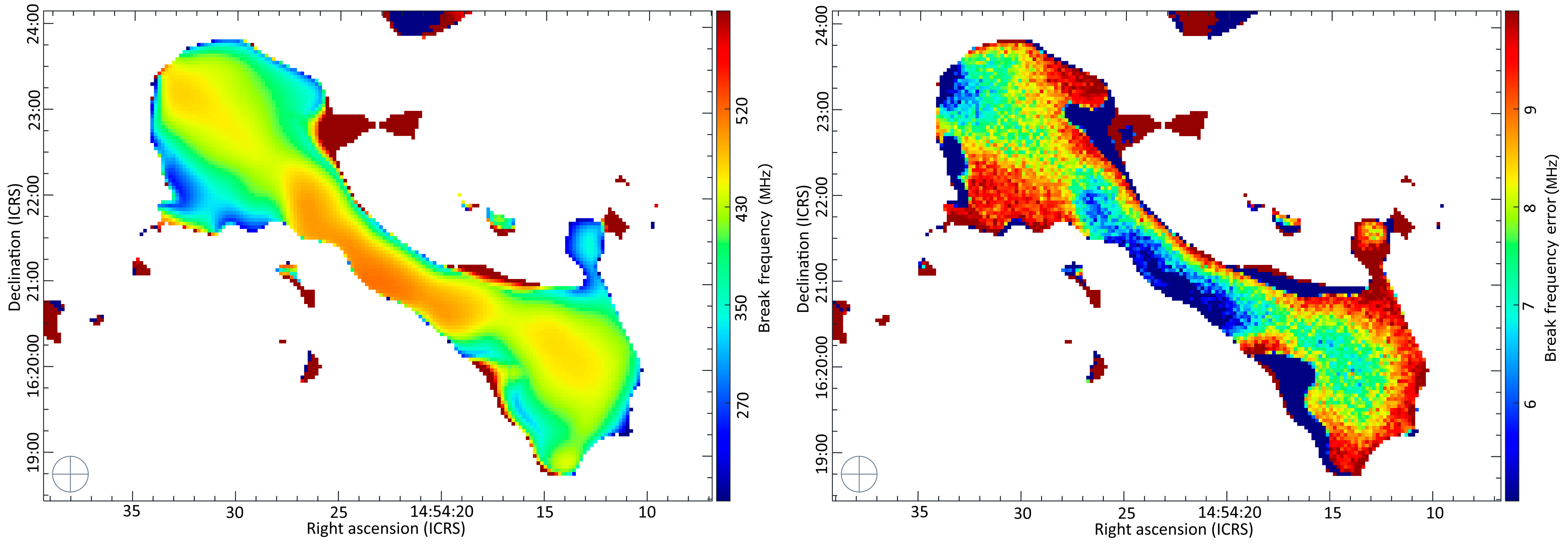}
\caption{On the left, the break frequency map of 3C\,306. The break frequency, $\nu_\mathrm{br}$, marks the point at which the synchrotron spectrum steepens due to radiative energy losses, with higher values of $\nu_\mathrm{br}$  tracing regions of more recent or ongoing particle acceleration. The map was derived from multi-frequency flux measurements using the following datasets: TGSS ADR (0.1475\,GHz), RACS (0.8875\,GHz and 1.3675\,GHz), and VLA (1.4899\,GHz and 4.8600\,GHz). On the right, the corresponding uncertainty map of $\nu_\mathrm{br}$, obtained through Monte Carlo perturbations of the input flux densities followed by repeated refitting using the synchrotron self-absorption plus spectral ageing (SSA+aging) model. Higher uncertainties appear in regions of lower signal-to-noise or where the spectral shape is less well constrained by the available frequency coverage.}
\label{fig:bf}
\end{figure*}

The statistical uncertainty on $\nu_\mathrm{br}$ was quantified using a Monte Carlo perturbation procedure. For each pixel, the observed intensities were perturbed with Gaussian noise at the level of $\sigma = \epsilon \, S_\nu$, where $\epsilon$ is the assumed relative flux calibration error (here $\epsilon = 0.05$). The perturbed datasets were re-fitted $N_{\rm MC}=100$ times, producing a distribution of $\nu_\mathrm{br}$ values. The $1\sigma$ uncertainty was then taken as the standard deviation of the recovered $\nu_\mathrm{br}$ distribution. This procedure naturally accounts for both measurement noise and the non-linear propagation of errors through the model fitting. The outcome of this procedure are two-dimensional maps of the break frequency $\nu_\mathrm{br}$ and its uncertainty $\sigma_{\nu_\mathrm{br}}$, as shown in \autoref{fig:bf}.

A physical interpretation of $\nu_\mathrm{br}$ can be obtained by connecting it to the synchrotron spectral age through \autoref{eq-tspec}. For a fixed value of $B$, lower values of $\nu_\mathrm{br}$ correspond to regions where the relativistic electron population has undergone longer radiative cooling timescales, while higher $\nu_\mathrm{br}$ values trace younger plasma, more recently injected or re-accelerated. Thus, the spatial distribution of $\nu_\mathrm{br}$ across 3C\,306 provides a useful indication of the relative age variations within the radio source. The break-frequency map show in \autoref{fig:bf} shows modest variations across 3C\,306, suggesting that the radiative ageing of the relativistic electron population progresses in a relatively homogeneous manner. A more detailed analysis of the spectral ageing timescale across 3C\,306 is provided in \autoref{subsec:SpectralAge}.

\subsection{Lobe-integrated SEDs and JP modelling}\label{subsec:JPmod}

 Given the relative uniformity of the break-frequency map (\autoref{fig:bf}), we performed an independent consistency check by fitting classical Jaffe--Perola (JP) \citep{1973A&A....26..423J} ageing models to lobe-integrated radio spectra. This analysis is entirely independent of the pixel-based modelling presented in the \autoref{section:ssa}, as it relies on integrated flux-density measurements compiled from the literature (see Tables~\ref{tab:sed_lobes}), rather than on the multi-frequency radio images used to derive the spatially resolved $\nu_{\rm break}$ maps.

 This complementary approach provides lobe-averaged estimates of the injection index and break frequency, which can be directly compared with the values inferred from the pixel-based analysis.

 \begin{table}
\centering
\caption{Radio flux density measurements of the northeastern (NE) and southwestern (SW) lobes of 3C\,306, extracted within circular regions of 35 arcsec radius.}
\label{tab:sed_lobes}
\begin{tabular}{ccc}
\hline
Frequency [GHz] & NE Flux density [Jy] & SW Flux density [Jy] \\
\hline
0.23100 & $3.17 \pm 0.06^{\dagger}$ & $2.31 \pm 0.05^{\dagger}$ \\
0.22300 & $2.82 \pm 0.16^{\dagger}$ & $2.28 \pm 0.13^{\dagger}$ \\
0.21600 & $2.99 \pm 0.16^{\dagger}$ & $2.08 \pm 0.13^{\dagger}$ \\
0.20800 & $2.84 \pm 0.18^{\dagger}$ & -- \\
0.20000 & $2.92 \pm 0.15^{\dagger}$ & $2.29 \pm 0.12^{\dagger}$ \\
0.19300 & $3.18 \pm 0.12^{\dagger}$ & $2.41 \pm 0.10^{\dagger}$ \\
0.18496 & $3.65 \pm 0.11^{\dagger}$ & $2.61 \pm 0.09^{\dagger}$ \\
0.17700 & $3.70 \pm 0.11^{\dagger}$ & $2.54 \pm 0.09^{\dagger}$ \\
0.17000 & $3.72 \pm 0.05^{\dagger}$ & $2.70 \pm 0.04^{\dagger}$ \\
0.16200 & $3.80 \pm 0.05^{\dagger}$ & $2.91 \pm 0.04^{\dagger}$ \\
0.15400 & $4.16 \pm 0.05^{\dagger}$ & $3.02 \pm 0.04^{\dagger}$ \\
0.14700 & $4.33 \pm 0.06^{\dagger}$ & $3.16 \pm 0.05^{\dagger}$ \\
0.13400 & $4.71 \pm 0.07^{\dagger}$ & $3.41 \pm 0.06^{\dagger}$ \\
0.12600 & $4.78 \pm 0.07^{\dagger}$ & $3.56 \pm 0.07^{\dagger}$ \\
0.11800 & $5.00 \pm 0.08^{\dagger}$ & $3.88 \pm 0.07^{\dagger}$ \\
0.11100 & $5.18 \pm 0.10^{\dagger}$ & $3.97 \pm 0.09^{\dagger}$ \\
0.10300 & $5.37 \pm 0.11^{\dagger}$ & $4.09 \pm 0.11^{\dagger}$ \\
0.09500 & $5.74 \pm 0.13^{\dagger}$ & $4.46 \pm 0.12^{\dagger}$ \\
0.08800 & $5.99 \pm 0.15^{\dagger}$ & $4.69 \pm 0.14^{\dagger}$ \\
0.08000 & $6.08 \pm 0.19^{\dagger}$ & $4.91 \pm 0.18^{\dagger}$ \\
\hline
0.40800 & $2.08 \pm 0.42^{\ddagger}$ & -- \\
1.40000 & $0.0332 \pm 0.0066^{\ddagger}$ & $0.0525 \pm 0.0105^{\ddagger}$ \\
\hline
\end{tabular}

\parbox{\columnwidth}{\vspace{1ex}
\textbf{Notes.} 
$^{\dagger}$ Measurements from \citet{2017MNRAS.464.1146H}. 
$^{\ddagger}$ Measurements from \citet{2010A&A...511A..53V}.}
\end{table}

For the northeastern (NE) lobe, the flux densities listed in \autoref{tab:sed_lobes} were extracted within a circular aperture of radius 35~arcsec centred at \(\mathrm{RA}=223.6250\), \(\mathrm{Dec}=16.3811\). The spectrum was modelled with the JP prescription, which assumes rapid isotropization of electron pitch angles in a locally uniform magnetic field, yielding an exponential steepening above a characteristic break frequency. The flux density at frequency \(\nu\) is written as

\begin{equation}
S(\nu) \;=\; S_0\,\left(\frac{\nu}{\nu_0}\right)^{-\alpha_{\rm inj}} \,
G\!\left(\frac{\nu}{\nu_{\rm break}}\right),
\end{equation}
\\where \(S_0\) is a normalization at \(\nu_0\), \(\alpha_{\rm inj}\) is the injection spectral index, \(\nu_{\rm break}\) is the radiative break frequency, and \(G(x)\) is the JP response function (the synchrotron kernel integrated over an isotropized electron population; see \citealt{1973A&A....26..423J} for the explicit form). 

\begin{figure}
    \centering
    \includegraphics[width=\columnwidth]{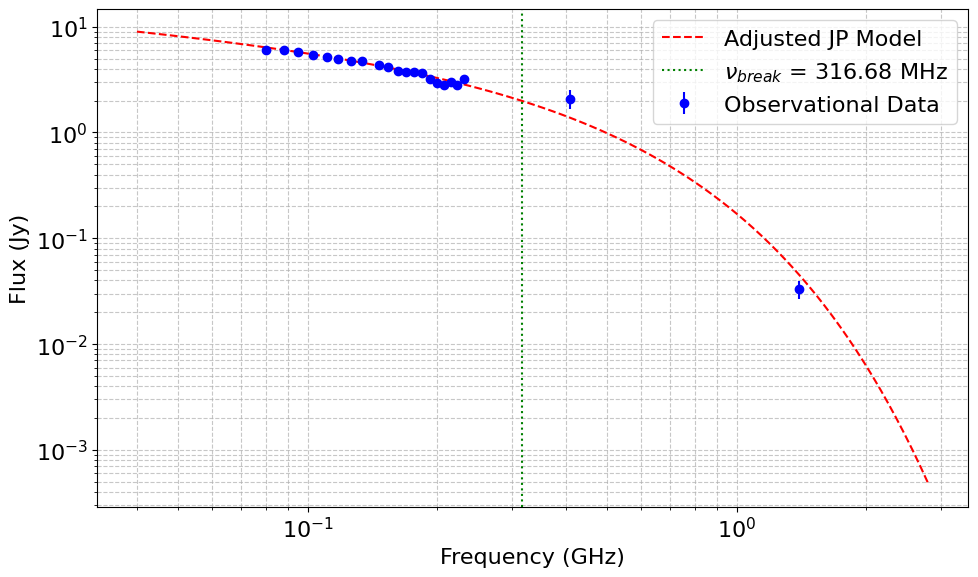}
    \caption{Spectral fit of the northeastern lobe of 3C\,306 using the Jaffe--Perola (JP) model. 
    Observed flux densities are shown as blue points with uncertainties, while the best-fit JP model is displayed as a dashed red line. 
    The vertical green line indicates the fitted break frequency.}
    \label{fig:jp_fit_se_lobe}
\end{figure}

We performed weighted non-linear least-squares fits using the tabulated uncertainties. The free parameters were \(S_0\), \(\alpha_{\rm inj}\), and \(\nu_{\rm break}\). For the NE lobe we obtain $S_0 = 2.71 \pm 0.49$ Jy, $\alpha_{\rm inj} = 0.10 \pm 0.06$, and $\nu_{\rm break} = 316.7 \pm 31.4$ MHz, with the best-fitting shown in \autoref{fig:jp_fit_se_lobe}. The break frequency obtained for the NE lobe from the JP fit is lower than the values inferred for the same region from the pixel-based map (\autoref{fig:bf}), where the characteristic frequencies are typically of the order of $\sim 440$ MHz. This difference is not unexpected, since the two approaches probe the spectrum in different ways. The JP fit is based on lobe-integrated flux densities compiled from the literature and therefore provides a single average spectral solution for the entire aperture, whereas the pixel-based analysis samples the local spectral curvature across the source using a different parametrization (Eq.~\ref{eq:ssa_aged}). In addition, the integrated spectrum has relatively sparse coverage between $\sim 0.2$ and $2$ GHz, precisely the range that is most relevant for constraining the spectral steepening. The derived $\nu_{\rm break}$ may also depend on aperture averaging, since regions with distinct spectral ages are combined into a single spectrum in the lobe-integrated approach. For these reasons, the NE JP value should be regarded as a complementary, aperture-averaged estimate rather than a direct one-to-one equivalent of the local break frequencies measured from the spatially resolved map.

\begin{figure}
    \centering
    \includegraphics[width=\columnwidth]{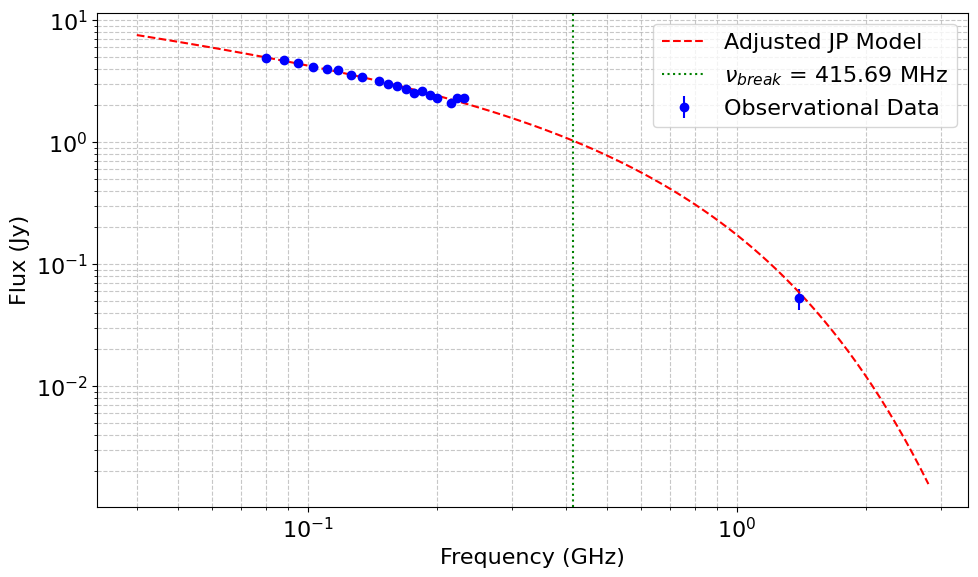}
    \caption{Spectral fit of the southwestern lobe of 3C\,306 using the Jaffe--Perola (JP) model. 
    Observed flux densities are shown as blue points with uncertainties, while the best-fit JP model is displayed as a dashed red line. 
    The vertical green line indicates the fitted break frequency.}
    \label{fig:jp_fit_sw_lobe}
\end{figure}

 We repeated the procedure for the southwestern (SW) lobe using the measurements in \autoref{tab:sed_lobes}, extracted within a 35~arcsec aperture centred at \(\mathrm{RA}=223.5583\), \(\mathrm{Dec}=16.3350\). The resulting JP fit, shown in \autoref{fig:jp_fit_sw_lobe}, provides a good description of the observed flux distribution, yielding best-fitting parameters of \(S_0 = 1.26 \pm 0.22\)~Jy, \(\alpha_{\rm inj} = 0.26 \pm 0.06\), and \(\nu_{\rm break} = 415.7 \pm 51.0\)~MHz.

Taken together, the JP modelling of both lobes indicates break frequencies of \(\nu_{\rm break} = 316.7 \pm 31.4\)~MHz (NE) and \(\nu_{\rm break} = 415.7 \pm 51.0\)~MHz (SW), with injection indices of \(\alpha_{\rm inj}=0.10 \pm 0.06\) and \(\alpha_{\rm inj}=0.26 \pm 0.06\), respectively. Despite these differences, both lobes exhibit comparable spectral curvature, consistent with a broadly similar radiative history. The discrepancy between the injection indices derived from the JP fits and the fixed value of $\alpha_{\rm thin}$ adopted in the pixel-based modelling primarily reflects the different assumptions and sensitivities of the two approaches. While the JP analysis provides an integrated, lobe-averaged estimate of the spectral properties, the spatially resolved modelling relies on a parametrized description of the local spectra, in which $\alpha_{\rm thin}$ is fixed to ensure stability of the fits. In this context, the JP modelling provides a global characterization of the spectral ageing that is in overall agreement with the trends inferred from the spatially resolved analysis (\autoref{fig:bf}), and therefore serves as an independent consistency check of the spectral modelling.

 \begin{figure*}
 \centering
 \includegraphics[width=\textwidth]{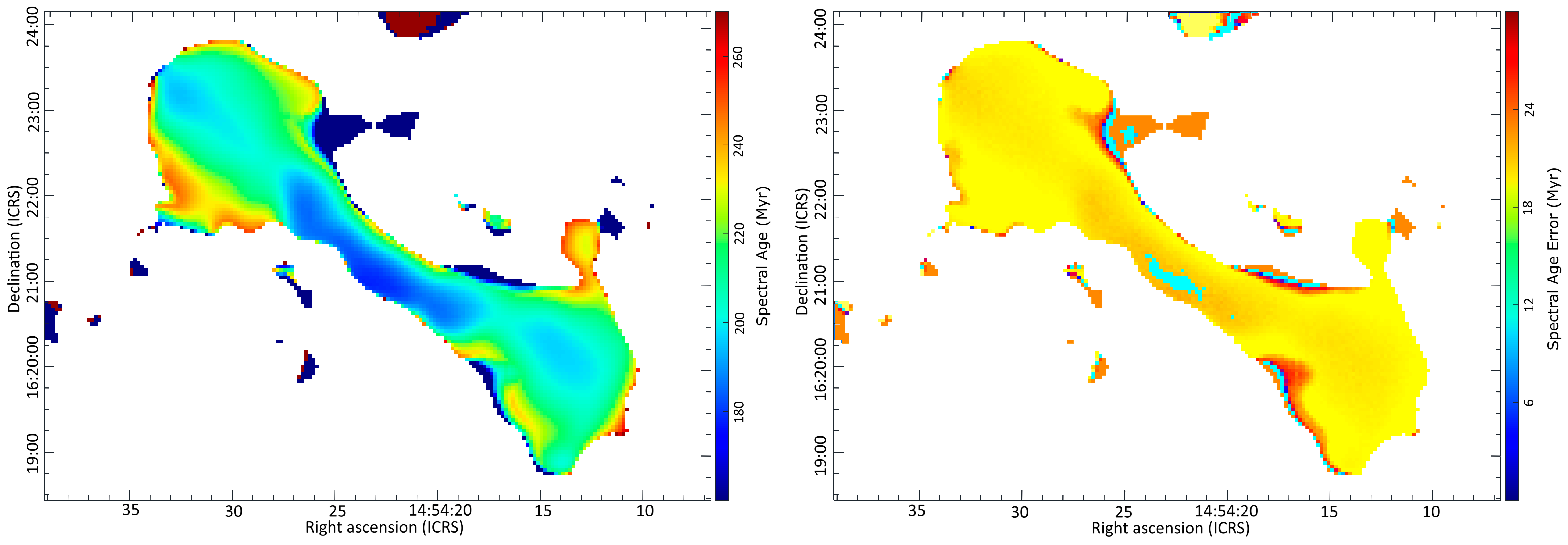} 
 \caption{Spectral age map ($t_\mathrm{spec}$) generated from the break frequency ($\nu_\mathrm{br}$) and considering $z = 0.0453$ and $B_\mathrm{eq} = 2.1\; \mu$G.}
 \label{fig:spectralage}
\end{figure*}

\subsection{Equivalent magnetic field estimates}\label{eqB}

 The radio power at frequency $\nu$, $P^{\mathrm{rad}}_{\nu}$, is calculated from the spectral luminosity $L_{\nu}$ using the flux density $S_{\nu}$ measured directly from the 1.40-GHz VLA continuum image. The flux was obtained by integrating the surface brightness over the full extent of the source emission, ensuring that both lobes and the bridge region are included. The radio power is then computed as

\begin{equation}
P^{\mathrm{rad}}_{\nu}=\nu L_{\nu}=\nu\frac{4\pi D_\mathrm{L}^2}{\left(1+z\right)^{1-\alpha}}  S_{\nu}
\end{equation}
\\where $D_L$ represents the luminosity distance to the source.  

 Additionally, uncertainties in the calculated radio power were estimated through standard error propagation, taking into account both flux density measurements and luminosity distance uncertainties \citep{2011AIPC.1381..117B}. From this procedure, we obtain a radio power at 1.4 GHz of 
$P^{\mathrm{rad}}_{1.4\mathrm{GHz}} = (2.92 \pm 0.14) \times 10^{40}$ erg s$^{-1}$. 
This value is broadly consistent with previous estimates in the literature, such as $8.81\times 10^{40}$ erg s$^{-1}$ at 1.1 GHz reported by \citet{2005A&A...440..775K}, and the $\sim 10^{41}$ erg s$^{-1}$ derived by \citet{2017MNRAS.466.4346M}.

Assuming that the radio emission arises from relativistic particles embedded in a uniformly distributed magnetic field under equipartition conditions, the magnetic field strength, $B_\mathrm{eq}$, can be estimated following classical formulations (e.g., \citealt{1964ocr..book.....G,1970ranp.book.....P,2005AN....326..414B}):

\begin{equation}
B_\mathrm{eq} = 9.6 \times 10^{-12} \left( \frac{(1+k)\,L_{1.4\mathrm{GHz}}}{V} \right)^{2/7},
\end{equation}
\\ where $L_{\nu}$ is the monochromatic radio luminosity at 1.4 GHz (in W Hz$^{-1}$), $k$ is the ratio between the energy in non-radiating particles and relativistic electrons (assumed to be unity), and $V$ is the volume of the radio-emitting region.

In this work, we assume an ellipsoidal geometry to estimate the emitting volume, corresponding to the region where the spectral index is consistent with the injection value ($\alpha \simeq \alpha_\mathrm{inj}$). Using the luminosity derived from the observed flux density and the estimated volume, we obtain an equipartition magnetic field of $B_\mathrm{eq} = 2.1 \pm 0.3\ \mu$G.

Such a value is consistent with magnetic fields of a few $\mu$G inferred in the lobes of other radio galaxies from IC/CMB studies. For instance, 3C~98 shows $B \approx 1.7~\mu\mathrm{G}$ \citep{2005ApJ...632..781I}; in Fornax~A one obtains $B_{\mathrm{IC}} \sim 2\text{--}3~\mu\mathrm{G}$ \citep{1995ApJ...449L.149F,1995ApJ...453L..13K}; and for samples of FR~II sources the typical fields lie close to $(0.3\text{--}1.3)\,B_{\mathrm{eq}}$, i.e. also within the $\mu$G range \citep{2005ApJ...626..733C}.

\subsection{Spectral Age}\label{subsec:SpectralAge}

The spectral age $t_\mathrm{spec}$ was calculated on a pixel-by-pixel basis using the break frequency values shown in \autoref{fig:bf} and \autoref{eq-tspec}. In this framework, $t_\mathrm{spec}$ acts as a radiative clock, tracing the time elapsed since the last significant acceleration of the relativistic electrons. For the calculations, we adopted the equipartition magnetic field $B_\mathrm{eq} = 2.1 \pm 0.3\ \mu$G for $B$ across 3C\,306 and the value of $B_\mathrm{CMB}$ at the source's redshift. 
The resulting map is shown in \autoref{fig:spectralage}, providing a spatially resolved estimate of the synchrotron lifetimes across the radio lobes.

The spectral age distribution indicates that most of the radio emission is characterized by $t_{\mathrm{spec}} \approx 220\,\mathrm{Myr}$. No sharp discontinuities are evident in the spectral age map. Instead, the distribution follows a smooth spatial trend, with comparatively lower spectral ages toward the central regions and higher values toward the outer parts of the radio lobes. In particular, the spectral ages increase gradually along the jet axis, reaching values of $\gtrsim 230\,\mathrm{Myr}$ at the lobe extremities, while regions closer to the core typically exhibit $t_{\mathrm{spec}} \lesssim 200\,\mathrm{Myr}$.

While this behaviour is qualitatively consistent with progressive radiative ageing of the electron population, it is important to emphasize that the derived spectral ages represent effective, beam-averaged quantities. The radio images were convolved to a common angular resolution of $25\arcsec$, which leads to spatial mixing of electron populations with different ages within each resolution element. As a consequence, localized regions of freshly accelerated plasma may not be resolved, and their contribution can be diluted in the observed spectrum. In addition, the frequency coverage of the available data, spanning up to 4.86\,GHz, limits the sensitivity to very high break frequencies. Since recently accelerated electrons are expected to exhibit $\nu_{\mathrm{br}}$ values beyond this range, the absence of very young spectral ages in the map does not necessarily imply a lack of ongoing particle acceleration, but rather reflects the limited ability of the data to constrain such high-frequency spectral curvature.

The uncertainties shown in the spectral age map are relatively small, reinforcing the internal consistency of the fitting procedure, although deviations in low surface-brightness regions should be interpreted with caution.

The overall smoothness of the spectral age distribution suggests a relatively coherent large-scale evolution of the electron population. However, strong conclusions regarding the temporal behaviour of the jet activity — such as the presence or absence of intermittent phases or multiple episodes of particle injection — cannot be robustly established from the spectral ageing analysis alone, given the combined effects of spatial resolution, spectral modelling assumptions, and frequency coverage limitations.

The derived radiative timescale of $\sim 220$ Myr is consistent with the order of magnitude expected for FR~II radio galaxies on similar spatial scales. Nevertheless, the absolute values of $t_\mathrm{spec}$ depend on the assumed magnetic field strength and on the adopted spectral model; departures from equipartition or the presence of additional physical processes, such as re-acceleration or mixing, can systematically affect the inferred ages \citep[e.g.,][]{2013MNRAS.435.3353H,2002ASSL..272..163F}. In this context, the spectral age map should be interpreted primarily as tracing relative variations in the radiative history of the source, rather than providing a direct measurement of its dynamical age.

\section{Numerical simulations}\label{sec:NumSim}

\subsection{Initial setup}\label{sec:IniSet}

The methodology adopted for the magnetohydrodynamic (MHD) simulations in this work is similar to that presented in \citealt{2024MNRAS.533.1341A} for their hydrodynamic simulations concerning the radio galaxy 3C\,338. 
The MHD simulations were conducted with the open-source code PLUTO\footnote{\url{http://plutocode.ph.unito.it/}} version 4.4 \citep{2007ApJS..170..228M}. In the MHD framework, PLUTO evolves numerically in space and time the mass density, $\rho$, velocity, $\emph{v}$, thermal pressure, $P$, and  magnetic field, $\emph{B}$, of a fluid through:

\begin{equation} \label{massconserv}
\frac{\partial \rho}{\partial t} + \nabla\cdot\left(\rho\emph{v}\right) = 0,
\end{equation}

\begin{equation} \label{momconservMHD}
\frac{\partial \left(\rho\emph{v}\right)}{\partial t} + \nabla\cdot\left[\rho\emph{v}\emph{v}^T - \frac{\emph{B}\emph{B}^T}{4\pi}+I\left(P+\frac{\vert\emph{B}\vert^2}{8\pi}\right)\right] = -\rho\nabla\Phi_\mathrm{tot},
\end{equation}

\begin{equation} \label{energconservMHD}
\frac{\partial E}{\partial t} + \nabla\cdot\left[\left(E+P+\frac{\vert\emph{B}\vert^2}{8\pi}\right)\emph{v}-\left(\emph{v}\cdot\emph{B}\right)\emph{B}\right] = - \rho\emph{v}\cdot\nabla\Phi_\mathrm{tot},
\end{equation}
\\where $I$ is the identity tensor of rank 3,  $\Phi_\mathrm{tot}$ is the dark matter (DM) gravitational potential and $E$ is the total energy density, defined as

\begin{equation} \label{totEMHD}
E = \frac{P}{\left(\Gamma-1\right)} + \frac{\rho\vert\emph{v}\vert^2}{2}+\frac{\vert\emph{B}\vert^2}{8\pi},
\end{equation}
\\where $\Gamma$ is the adiabatic index of the fluid (= 5/3 in this work) and $\emph{B}$ respecting the ideal induction equation

\begin{equation} \label{inductioneq}
\frac{\partial \emph{B}}{\partial t} = \nabla\times \left(\emph{v}\times\emph{B}\right).
\end{equation}

\begin{figure}
    \centering
    \begin{subfigure}{\columnwidth}
       \centering         \includegraphics[width=\columnwidth]{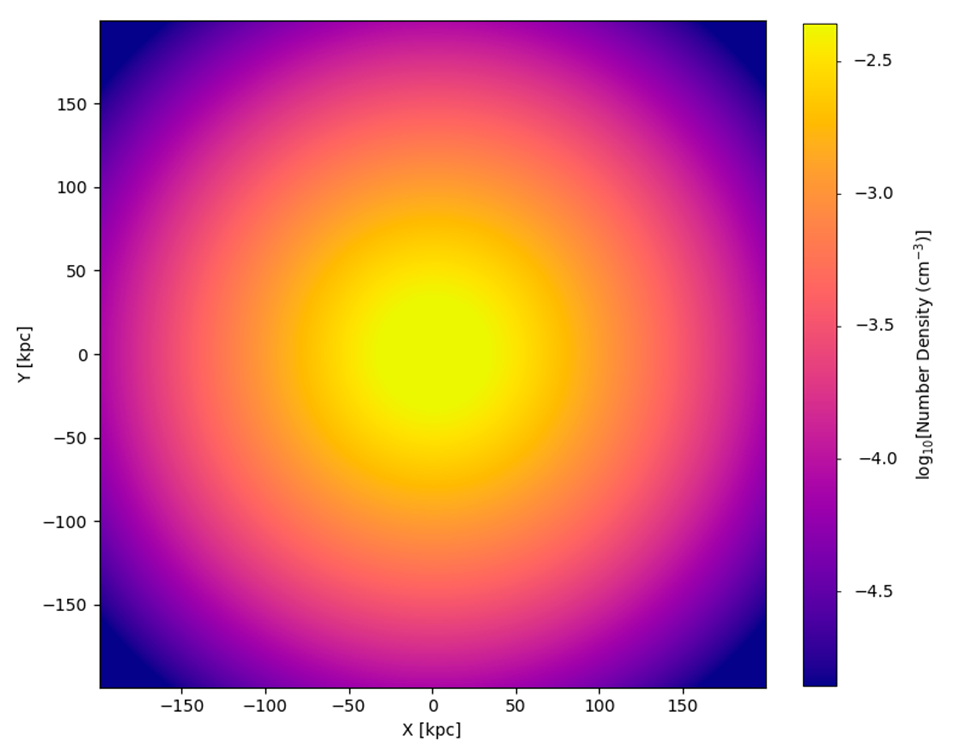}
    \end{subfigure}
     \hfill
     \begin{subfigure}{\columnwidth}
       \centering         
       \includegraphics[width=0.98\columnwidth]
       {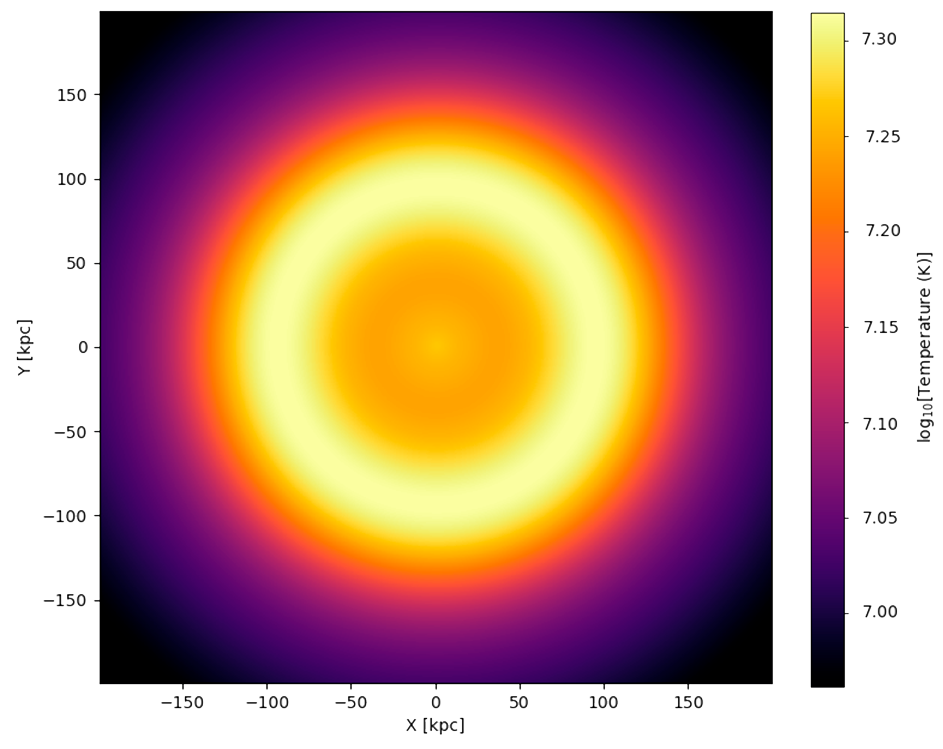}
         \label{fig:PKS0502_flux_figure}
     \end{subfigure}
     \hfill
     \caption{Initial spatial distribution of the number density (upper panel) and temperature (lower panel)
     on the plane $z=0$ adopted in our MHD simulations.}
    \label{fig:rho_T_cutxy0}
\end{figure}

The calculation of the numerical fluxes through adjacent cell interfaces were performed from the Harten, Lax, Van Leer approximate Riemann Solver for middle contact discontinuities (HLLC; \citealt{1994ShWav...4...25T, 2005JCoPh.203..344L}). The temporal evolution of $\rho$, $\rho\emph{v}$, $E+\rho\Phi_\mathrm{tot}$ 
and $B$
were done via the third-order Runge–Kutta method, while a piecewise parabolic method (PPM) interpolation was applied to reconstruct the primitive variables in each time step of our simulation. The solenoid condition ($\nabla\cdot\emph{B}=0$) in our MHD simulations was controlled by the eight-wave formalism \citep{1994arsm.rept.....P, 1999JCoPh.154..284P}.

The computational domain is cubic, having a physical length of 400 kpc along each Cartesian axis.
The numerical resolution is 1 kpc cell$^{-1}$ for our simulations that adopted a jet inlet radius of 4 kpc (4 numerical cells per jet inlet radius). The numerical frontiers of the computational domain were set as selective boundary condition \citep{2023ApJ...944...11C} with a threshold speed of 0.5 km s$^{-1}$ \citep{2024MNRAS.533.1341A}. 
Under this prescription, if the velocity component normal to the boundary is directed outwards and exceeds the adopted threshold, the normal gradient of any primitive variable across the boundary is set to zero (e.g., \citealt{2007ApJS..170..228M}). Otherwise, the velocity component normal to the boundary is set to zero.

The simulations were run during $\approx$250 Myr, compatible with the spectral ages of the lobes of 3C\,306 estimated in \autoref{subsec:SpectralAge}. Simulation outputs (spatial distributions of mass density, temperature, thermal pressure, 3D velocity and magnetic field vectors) were generated in time steps of 50 Myr. We assumed an initial static ($\emph{v}=\emph{0}$) and non-magnetised ($\emph{B}=\emph{0}$) environment with a number density and temperature radial profiles described in section \autoref{sec:xraydata}. We show in \autoref{fig:rho_T_cutxy0} the initial gas density and temperature distributions on the plane $z=0$ adopted in this work. The number density is about seventy times higher in the central region in relation to the outer parts. A non-relativistic jet/counter-jet is introduced in such environment, as described in the next section.

\begin{figure}
	\includegraphics[scale=0.58]{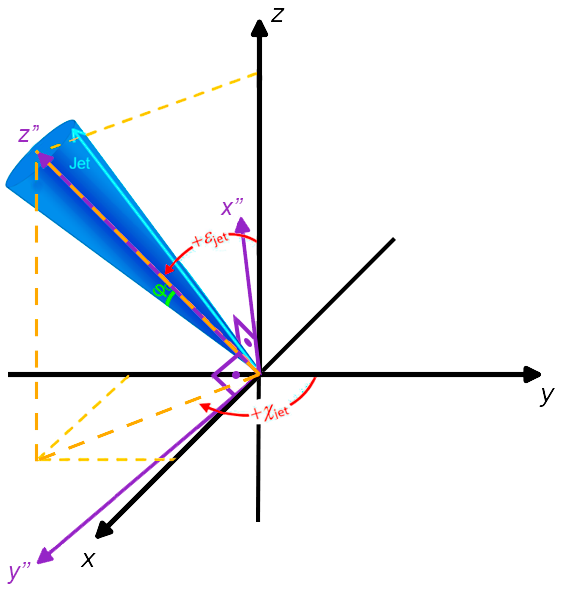}
	\centering
    \caption{Schematic representation of the jet orientation with respect to the $Oxyz$ reference frame (fixed to the computational domain of our simulations) and the $Ox''y''z''$ frame (rotated through the angles $\epsilon_\mathrm{jet}$ and $\chi_\mathrm{jet}$). The blue conical surface represents the opening angle of a precessing jet, corresponding to the case $\varphi_0 > 0$. The light-blue arrow indicates the instantaneous jet direction, while the dashed line marks the mean jet direction. } 
    
    \label{fig:epsilonchijet}
\end{figure}

\subsection{Jet/counter-jet injection}\label{sec:JetInj}

Similar to \citet{2017A&A...606A..57R} and \citet{2024MNRAS.533.1341A}, the injection of a pair of jets was performed by selecting two groups of adjacent cells confined within a cylindrical radius $r_\mathrm{jet}$ where the jet numerical density, $n_\mathrm{jet}$, jet pressure, $P_\mathrm {jet}$, jet speed $v''_\mathrm{jet}$, and a toroidal magnetic field, $ \emph{B}^{''}_\mathrm{jet}$, are imposed. 
For counter-jet injection, the signs of $v''_\mathrm{jet}$ and $ \emph{B}^{''}_\mathrm{jet}$ were reversed (e.g., \citealt{2017A&A...606A..57R}).

In our MHD simulations, the Cartesian $z$-axis is aligned along the line of sight towards the observer, while the $x$- and $y$-axes are directed westwards and northwards, respectively. The jet (and counter-jet) orientation in our simulations is defined through the angles $\varepsilon_\mathrm{jet}$, measured positively from the $z$-axis, and $\chi_\mathrm{jet}$, measured positively from the $y$-axis toward the $x$-axis, as shown in \autoref{fig:epsilonchijet}. Note that when both angles are set to zero, the jet axis (for a non-precessing jet) or the precession axis (in the case of jet precession) aligns with the $z$-axis, with the jet and counter-jet propagating towards $+z$ and $-z$, respectively\footnote{The jet precession cone axis (or the propagation direction in the non-precessing case) was chosen to differ from the computational grid $z$-axis for two primary reasons. First, it simplifies the construction of our synthetic synchrotron maps by keeping the line of sight—and thus the emissivity integration path—strictly fixed along the grid $z$-axis. Second, it accommodates time-evolving displacements of the jet inlet region, whether simple constant motion (as assumed in \autoref{pmc}) or more complex orbital geometries such as those considered in \citet{2024MNRAS.533.1341A}.}.

In the case of a precessing jet, the direction of the jet injection into the computational domain varies with time, forming an apparent precession cone with semi-aperture angle $\varphi_0$. In this case, $\varepsilon_\mathrm{jet}$ represents the angle between the precession cone axis and the $z$-axis, while $\chi_\mathrm{jet}$ is the position angle of the projection of this cone axis onto the $xy$-plane (measured positively from $y$ to $x$-axis).
The instantaneous orientation of the velocity vector evolves with time as

\begin{equation}
\label{vjxvjyvjzdprime}
\begin{cases}
    v''_{\mathrm{jet},x''} = v''_\mathrm{jet}\sin\varphi_0\sin\left(\varsigma2\pi\Delta\tau\right)\\
    v''_{\mathrm{jet},y''} = v''_\mathrm{jet}\sin\varphi_0\cos\left(\varsigma2\pi\Delta\tau\right)\\
    v''_{\mathrm{jet},z''} = v''_\mathrm{jet}\cos\varphi_0
\end{cases}
\end{equation}
\\where $\varsigma$ is the precession direction ($\varsigma=1$ for $+y''$ to $+x''$ and $\varsigma=-1$ for $+y''$ to $-x''$ sense of precession), $\Delta\tau=\tau-\tau_\mathrm{ph}=(t-t_\mathrm{ph})/P_\mathrm{prec,s}$, $P_\mathrm{prec,s}$ is the jet precession period at source's reference frame, and $\tau_\mathrm{ph}$ is a phase representing the normalized time when the jet/counter-jet crosses the $y''z''$-plane.

Their respective projections onto the $Oxyz$ reference frame are given as

\begin{equation}
\label{vjxvjyvjz}
\begin{cases}
    v_{\mathrm{jet},x} = v''_{\mathrm{jet},x''}\cos\chi_\mathrm{jet} + \left(v''_{\mathrm{jet},y''}\cos\varepsilon_\mathrm{jet} + v''_{\mathrm{jet},z''}\sin\varepsilon_\mathrm{jet}\right)\sin\chi_\mathrm{jet}\\
    v_{\mathrm{jet},y} = -v''_{\mathrm{jet},x''}\sin\chi_\mathrm{jet} + \left(v''_{\mathrm{jet},y''}\cos\varepsilon_\mathrm{jet} + v''_{\mathrm{jet},z''}\sin\varepsilon_\mathrm{jet}\right)\cos\chi_\mathrm{jet}\\
    v_{\mathrm{jet},z} = -v''_{\mathrm{jet},y''}\sin\varepsilon_\mathrm{jet} + v''_{\mathrm{jet},z''}\cos\varepsilon_\mathrm{jet}
\end{cases}
\end{equation}
\\reducing to 
Equation 15 in \citet{2024MNRAS.533.1341A} for a non-precessing jet ($v''_{\mathrm{jet},x''}=v''_{\mathrm{jet},y''}=0$ and $v''_{\mathrm{jet},z''}=v''_\mathrm{jet}$).

As mentioned previously, we have assumed a purely toroidal configuration for the magnetic field in the inlet region, as adopted in some previous relativistic and non-relativistic MHD simulations of radio galaxies (e.g., \citealt{2014MNRAS.443.1482H, 2016MNRAS.461.2025E, 2017A&A...606A..57R, 2022A&A...662A...5G}). In the case of a precessing jet, the magnetic field inside $r_\mathrm{jet}$ is defined in terms of its Cartesian components as

\begin{equation}
\label{BjxBjyBjz}
\begin{cases}
    B_{\mathrm{jet},x} = B''_{\mathrm{jet},x''}\cos\chi_\mathrm{jet} + \left(
B''_{\mathrm{jet},y''}\cos\varepsilon_{\mathrm{jet}}
+
B''_{\mathrm{jet},z''}\sin\varepsilon_{\mathrm{jet}}
\right)\sin\chi_{\mathrm{jet}},\\
    B_{\mathrm{jet},y} = -B''_{\mathrm{jet},x''}\sin\chi_\mathrm{jet} + \left(
B''_{\mathrm{jet},y''}\cos\varepsilon_{\mathrm{jet}}
+
B''_{\mathrm{jet},z''}\sin\varepsilon_{\mathrm{jet}}
\right)
\cos\chi_{\mathrm{jet}},\\
    B_{\mathrm{jet},z} = -B''_{\mathrm{jet},y''}\sin\varepsilon_\mathrm{jet} +
B''_{\mathrm{jet},z''}\cos\varepsilon_{\mathrm{jet}}.
\end{cases}
\end{equation}
\\where

\begin{equation}
\label{BjxBjyBjzdprime}
\begin{cases}
    B''_{\mathrm{jet},x''} = \left(\frac{B''_\mathrm{0,jet}}{r_\mathrm{jet}}\right)\left[y\cos\left(\varsigma2\pi\Delta\tau\right)-x\cos\varphi_0\sin\left(\varsigma2\pi\Delta\tau\right)\right]\\
    B''_{\mathrm{jet},y''} = -\left(\frac{B''_\mathrm{0,jet}}{r_\mathrm{jet}}\right)\left[y\sin\left(\varsigma2\pi\Delta\tau\right)+x\cos\varphi_0\cos\left(\varsigma2\pi\Delta\tau\right)\right]\\
    B''_{\mathrm{jet},z''} = \left(\frac{B''_\mathrm{0,jet}}{r_\mathrm{jet}}\right)x\sin\varphi_0
\end{cases}
\end{equation}
\\where $B''_\mathrm{0,jet}$ is the magnetic field strength at $r=r_\mathrm{jet}$ For a non-precessing jet, \autoref{BjxBjyBjzdprime} simplifies to  $B''_{\mathrm{jet},x''}=B''_\mathrm{0,jet}(y/r_\mathrm{jet})$, $B''_{\mathrm{jet},y''}=-B''_\mathrm{0,jet}(x/r_\mathrm{jet})$, and $B''_{\mathrm{jet},z''}=0$.

\begin{table*}
\centering
\caption{Jet parameters adopted in the simulations. 
The jet power $L_{\rm jet}$ is computed as the sum of kinetic, thermal and magnetic contributions (see text). 
Uncertainties are dominated by the choice of $B_{\rm jet}$ and the plasma composition; values should therefore be regarded as order-of-magnitude estimates.}
\label{tab:jetparams}
\begin{tabular}{lccccccccc}
\hline
Model & $r_{\rm jet}$ & $n_{\rm jet}$ & $v_{\rm jet}$ & $T_{\rm jet}$ & $B_{\rm jet}$ & Precession Period & $\eta_0$ & $v_{\rm src}$ & $L_{\rm jet}$ \\
      & [kpc] & [cm$^{-3}$] & [km\,s$^{-1}$] & [K]  & [$\mu$G] & [Myr] & [degrees] & [km\,s$^{-1}$] & [erg\,s$^{-1}$] \\
\hline
Pure Precession & 4.0 & $1.1\times10^{-5}$ & $2.998\times10^{4}$ & $4.51\times10^{8}$ & 2.1 & 250 & 240$^{\circ}$ & -- & $1.1\times10^{44}$ \\
Pure galactic motion     & 4.0 & $1.1\times10^{-5}$ & $2.998\times10^{4}$ & $4.51\times10^{8}$ & 2.1 & -- & -- & 75.2 (SE) & $1.1\times10^{44}$ \\
Composite       & 4.0 & $1.1\times10^{-5}$ & $2.998\times10^{4}$ & $4.51\times10^{8}$ & 2.1 & 250 & 234$^{\circ}$ & 75.2 (SE) & $1.1\times10^{44}$ \\
\hline
\end{tabular}
\end{table*}

\section{Results}\label{sec:SimRes}

We investigate three physically motivated MHD scenarios for 3C\,306: (i) pure jet precession, (ii) pure source motion relative to the ICM, and (iii) a composite model combining both effects. All simulations were evolved for a total time of 250 Myr, allowing the system to develop large-scale structures comparable to the observed radio morphology of 3C\,306. The jet power, $L_{\rm jet}$, was assumed to be time constant and defined as (e.g., \citealt{2023MNRAS.522.2850C}):

\begin{equation} \label{jetpowerMHD}
L_\mathrm{jet} = \left(1+\frac{u_\mathrm{th,jet}}{u_\mathrm{kin,jet}}+\frac{u_\mathrm{B,jet}}{u_\mathrm{kin,jet}}\right)\frac{1}{2}\rho_\mathrm{jet}(v''_\mathrm{jet})^3A_\mathrm{jet}.
\end{equation}
\\where $\rho_\mathrm{jet}$ is the jet mass density and $A_\mathrm{jet}$ is the jet cross-sectional area. The terms $u_\mathrm{th,jet}=\Gamma(\Gamma-1)^{-1}P_\mathrm{jet}$, $u_\mathrm{kin,jet}=1/2\rho_\mathrm{jet}(v''_\mathrm{jet})^2$, and $u_\mathrm{B,jet}=(B^{''}_\mathrm{jet,0})^2/(8\pi)$ correspond to the thermal, kinetic, and magnetic energy densities, respectively, with $\Gamma$ being the adiabatic index ($=5/3$ in this work).

The jet power was independently constrained from observations. To translate the radio emission into a mechanical jet power estimate, we employed the empirical cavity--radio scaling relation of \citet{2010ApJ...720.1066C}:

\begin{equation}
    L_{\rm jet} \;\approx\; 5.8 \times 10^{43}
    \left( \frac{P^{\mathrm{rad}}_{1.4\mathrm{GHz}}}{10^{40}\ \mathrm{erg\ s^{-1}}} \right)^{0.70}
    \ \mathrm{erg\ s^{-1}}.
\end{equation}

From the measured flux density (see \autoref{eqB}), we obtained a radio power of $P^{\mathrm{rad}}_{1.4\mathrm{GHz}} = (2.92 \pm 0.14) \times 10^{40}$ erg s$^{-1}$. Substituting this value into the scaling relation yields a mechanical jet power of $L_{\rm jet} \simeq 1.2 \times 10^{44}$ erg s$^{-1}$, with an intrinsic uncertainty of about 0.7 dex associated with the cavity--radio scaling relation. To reproduce this power in the simulations, the jet density was adjusted to $n_{\rm jet} \sim 10^{-5}\,\mathrm{cm^{-3}}$. 

At the injection point, the ratio of the jet thermal pressure to the ambient pressure is less than unity ($P_{\mathrm{jet}}/P_{\mathrm{amb}} \simeq 0.08$). For models incorporating source motion, the jet inlet region follows the prescribed orbital trajectory through the ICM (see \autoref{pmc} and \autoref{composite}), causing this pressure ratio to vary from approximately 0.08 to 0.10 along the orbit.

The main physical parameters adopted in the simulations are summarized in \autoref{tab:jetparams}.

\begin{figure}
\centering
\includegraphics[width=0.48\textwidth]{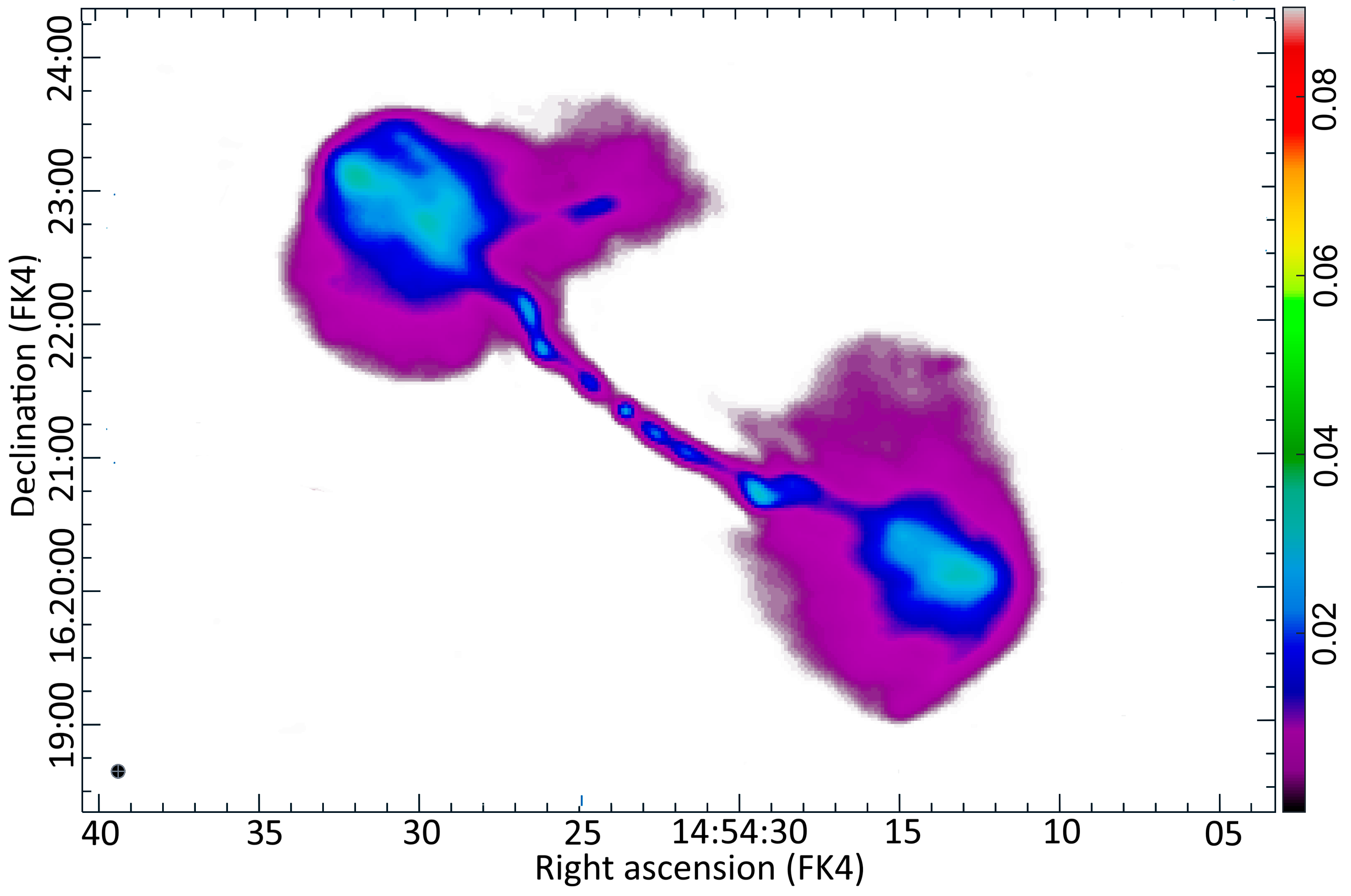}
\caption{
LoTSS DR3 144 MHz reference image of 3C\,306 \citep{2026A&A...707A.198S} used for the morphological comparison with the synthetic synchrotron maps. The colour scale represents the radio surface brightness in units of Jy beam$^{-1}$, while the black ellipse in the lower-left corner indicates the FWHM of the synthesised CLEAN beam.
}
\label{fig:ic4516_lotss_ref}
\end{figure}

To compute the synthetic synchrotron intensity maps, we assume that the radio emission in the optically thin regime is generated by a relativistic electron population with a power-law energy distribution, $N(E)\propto E^{-p}$, where $p = 2\alpha_{\rm thin}+1$. We have adopted $\alpha_{\rm thin}=0.625$, which corresponds to $p=2.25$. Following standard synchrotron formalism for a power-law electron distribution (e.g., \citealt{1970ranp.book.....P, 1979rpa..book.....R}), the local emissivity, $j_\nu$, in each computational cell is parameterised as

\begin{equation}\label{eq:synemiss}
j_{\nu}\propto\delta^{(p+3)/2} P B^{(p+1)/2}\nu^{-\alpha_{\rm thin}},
\end{equation}
\\ where $\delta$ is the relativistic Doppler boosting factor\footnote{Given the low jet velocity adopted in our numerical simulations, $\delta\approx1$.}.

The synthetic synchrotron intensity is then obtained by integrating $j_{\nu}$ pixel by pixel along the line of sight ($z$-axis). To allow a direct morphological comparison with the observational data, the resulting synthetic maps are convolved with a circular Gaussian beam matched to the angular resolution of the reference image (6 arcsec or 5.271 kpc at the redshift of 3C\,306). This reference image was obtained from the LOFAR Two-metre Sky Survey Data Release 3 (LoTSS DR3; \citealt{2026A&A...707A.198S}) and is shown in \autoref{fig:ic4516_lotss_ref}.

The synchrotron emissivity computed using \autoref{eq:synemiss} depends on the thermal pressure and magnetic field strength of the MHD flow. Consequently, recently injected jet material and freshly shocked regions, which generally retain higher pressures and stronger magnetic fields, tend to appear systematically brighter in the synthetic maps. Because this prescription does not explicitly account for spectral ageing, radiative losses, or the transport history of the relativistic electron population, it should be regarded as a simplified model aimed at capturing large-scale synchrotron morphology rather than fine brightness contrasts between young and aged emission regions.

\begin{figure}
\centering
\includegraphics[width=0.47\textwidth]{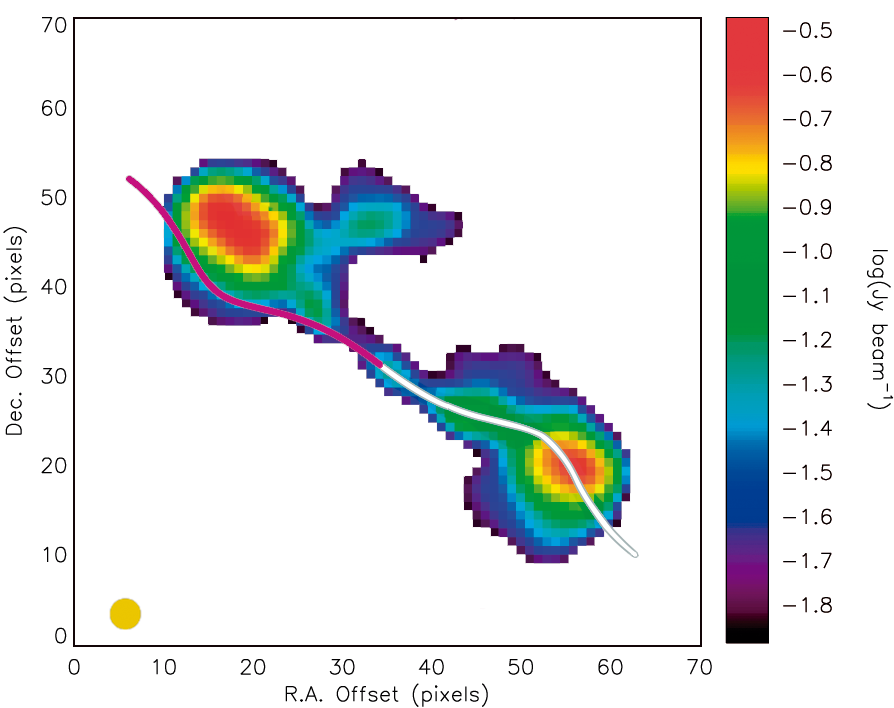}
\caption{Superposition of the modelled precession helices of the jet (white) and counter-jet (purple) over an interferometric image of 3C$\,$306 at 1.4 GHz. The curves represent an 250 Myr precession period, adopting the parameters $\tau_{0,s} = 0.49$, $\phi_0 = 50^{\circ}$, $\varphi_0 = 5^{\circ}$, $\eta_0 = 240^{\circ}$, $\iota = 1$, and $\beta = 0.0034$. The yellow ellipse in the lower-left corner indicates the FWHM of the synthesised CLEAN beam.}
\label{result_prec_model1}
\end{figure}

\subsection{Pure precession case}\label{sec:PurePrecMod}

The large-scale radio morphology of 3C$\,$306 presents distorted substructures (see \autoref{sec:radiodata}) that may naturally arise if the jet axis changes its orientation over time. 
To explore this possibility, we first examine the case where the observed structure is driven solely by jet/counter-jet precession.

\subsubsection{Kinematic jet precession model}\label{sec:PrecMod}

Following \citet{10.1111/j.1365-2966.2009.15361.x}, the jet and counter-jet of 3C\,306 are assumed to consist of plasma elements ejected with constant velocity $\beta c$, where $c$ is the speed of light and $0 \leq \beta < 1$. Due to precession, these elements are launched along directions that trace the surface of a cone with semi-opening angle $\varphi_0$, whose axis is inclined by an angle $\phi_0$ with respect to the line of sight. Precession leads to a periodic reorientation of the jet direction, with period $P_\mathrm{prec,obs}$ measured in the observer's frame. The successive ejection of plasma elements results in a helical structure around the projection of the precession cone axis onto the plane of the sky, forming a position angle $\eta_0$ measured from North to East. The remaining free parameters of the model are the precession phase, $\tau_{0,s}$, and the precession direction, $\iota$ ($\iota = 1$ for clockwise and $\iota = -1$ for counter-clockwise sense of precession).

The jet precession parameters were constrained qualitatively, without a formal statistical fitting procedure, using the 1.4 GHz radio map of 3C\,306 as the primary observational reference. It is shown in \autoref{result_prec_model1} the 1.4-GHz image of 3C\,306 overlaid with the precession helices generated by a model with $\beta = 0.0034$, $P_\mathrm{prec,obs} = 250$ Myr, $\phi_0 = 50^{\circ}$, $\varphi_0 = 5^{\circ}$, $\eta_0 = 240^{\circ}$, $\iota = 1$, and $\tau_{0,s} = 0.49$.
The jet and counter-jet of 3C\,306 are represented by white and purple curves, respectively. 
The outermost portions of the precession helices correspond to plasma elements ejected approximately 250 Myr ago. Note that the value of  $\beta \cos\phi_0$ implied by such precession model ($\sim 0.064$) is not so far from the value of $\sim 0.052$ derived from the arm-length ratio using the 0.89-GHz RACS observations of 3C\,306.

While this particular precession model provides a reasonable description of the overall jet morphology in the south-western structure of 3C\,306, it clearly fails to reproduce the source's counter-jet, departing systematically from the brightest components in the north-eastern structure. Testing alternative combinations of the model parameters also failed to simultaneously reproduce both the jet and the counter-jet. Even though these findings suggest that jet precession alone is not the most suitable scenario for 3C\,306, we chose to test this jet precession model in the framework of a MHD simulation, as discussed in the next section.

\subsubsection{MHD simulation of a precessing jet}

A MHD simulation of a jet/counter-jet under precession was performed to test the kinematic precession model presented in last section. 
The jet is turned on at $t = 0$ and remains active for 250 Myr. The remaining jet precession parameters assumed in the simulation were 
$P_\mathrm{prec,s}=250$ Myr\footnote{Given the relative low redshift of 3C\,306, we have ignored the cosmological term $(1+z)$ present in the relationship between the precession periods measured at the observer's and source's reference frames.}, $\varepsilon_\mathrm{jet}=50^\circ$, $\varphi_0=5^\circ$, $\chi_\mathrm{jet}=360^\circ-\eta_0=120^\circ$, $\varsigma=-\iota=-1$, and $\tau_\mathrm{ph}=\tau_\mathrm{0,s}+\iota/4=0.74$\footnote{Due to differences in the reference frames adopted in \citet{10.1111/j.1365-2966.2009.15361.x} and in this work, the last three quantities differ from each other, and their relationship is given explicitly in the text.}.
 As the density contrast between the jet and the surrounding environment is relatively high $n_0/n_\mathrm{jet}\sim 1000$, jet must interact strongly with the ICM, leading to significant deceleration as it propagates. The predicted bow-shock speed, $v_\mathrm{bs}$, under the highly supersonic flow condition can be estimated from (e.g., \citealt{1993ApJ...410..686D}):

\begin{equation} \label{eq:vbs}
v_\mathrm{bs} \approx \frac{v_\mathrm{jet}}{1+\left(r_\mathrm{h}/r_\mathrm{jet}\right)\left(\rho_\mathrm{amb}/\rho_\mathrm{jet}\right)^{1/2}},
\end{equation}
\\where $r_\mathrm{h}$ is the bow-shock (jet head) radius and $\rho_\mathrm{amb}$ is the mass density of the pre-shocked ambient.

\begin{figure*}
\centering
\includegraphics[width=\textwidth]{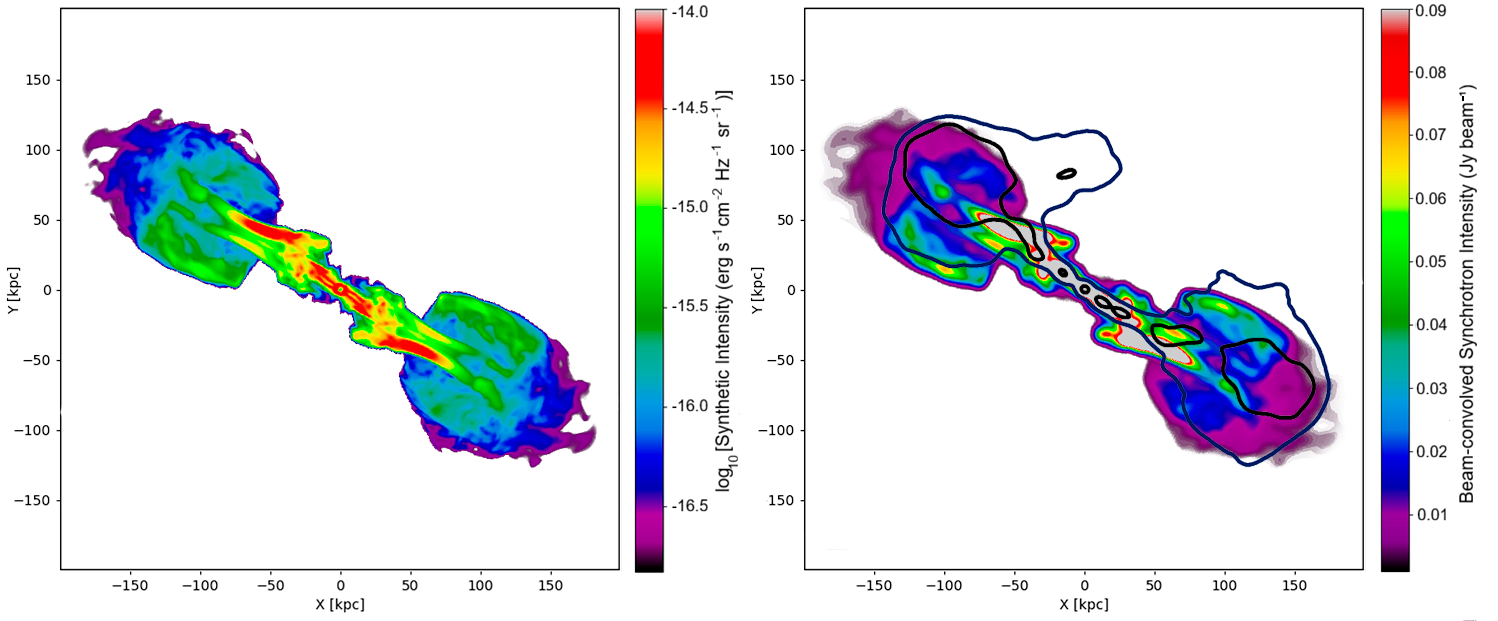}
\caption{Synthetic three-dimensional synchrotron intensity maps for the pure precession Case. In the left panel, the projected synchrotron intensity distribution is shown. In the right panel, the corresponding map convolved with a $6.0\arcsec \times 6.0\arcsec$ Gaussian beam and sampled at $0.6702\arcsec$ pixel$^{-1}$, matching the LoTSS DR3 144-MHz observation, is overlaid with observational contours at 0.005 Jy\,beam$^{-1}$ (dark blue) and 0.01 Jy\,beam$^{-1}$ (black).
}
\label{fig:mhdresult1}
\end{figure*}

\begin{figure}
\centering
\includegraphics[width=0.47\textwidth]{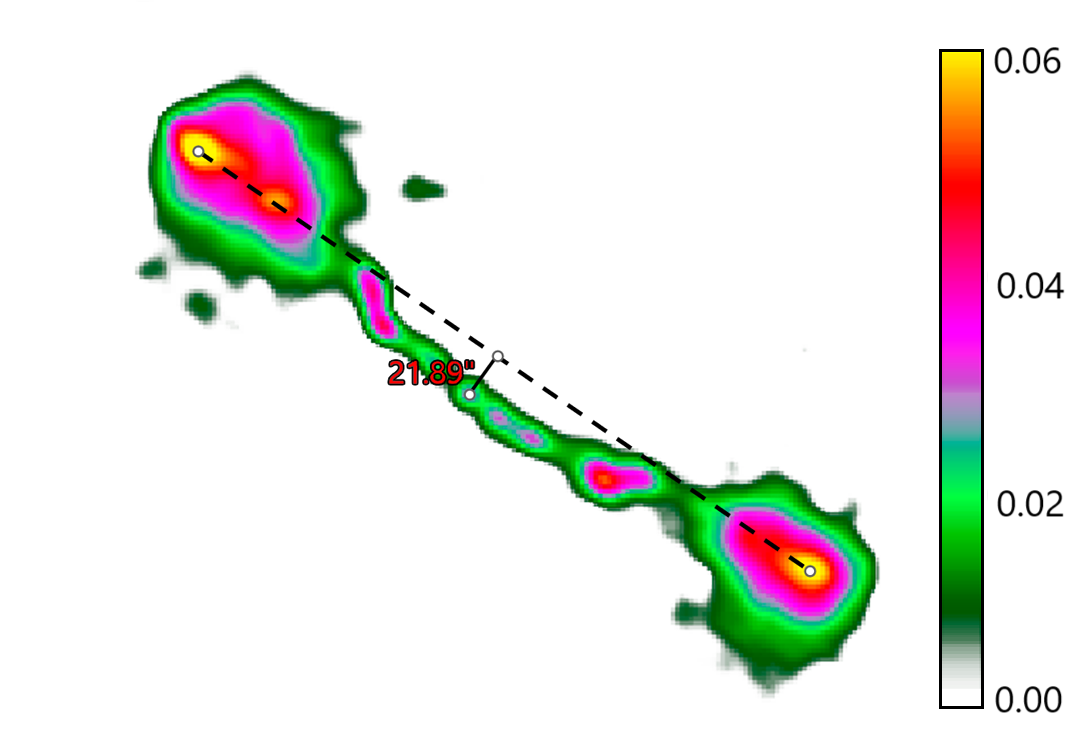}
\caption{Estimation of the displacement relative to the longitudinal direction defined by the brightest points of the lobes, superimposed on RACS low-I observations.}
\label{shift}
\end{figure}

Assuming a bow-shock radius comparable to the jet radius ($r_\mathrm{h}/r_\mathrm{jet} \approx 1$) and ambient densities typical of the jet inlet region, \autoref{eq:vbs} yields $v_\mathrm{jet} \approx 0.1c$ for an assumed $v_\mathrm{bs} = 0.0034c$. Consequently, the parameter $\beta$ used in the kinematic precession model (\autoref{sec:PrecMod}) should be interpreted as an effective (mean) jet velocity, reflecting the interaction between the jet and the ICM during its propagation.

The synthetic synchrotron map obtained for the pure precession model is shown in \autoref{fig:mhdresult1}. Compared with the baseline control simulation (see \autoref{app:baseline_model}), the emitting bridge connecting the two lobes remains broadly similar. This behaviour is consistent with the adopted precession geometry (see \autoref{result_prec_model1}), since the modelled jet and counter-jet trajectories remain relatively close to the mean source axis across much of the bridge, with larger departures occurring towards the outer regions. Consequently, the effects of precession become more apparent at larger distances from the nucleus, where the radio structure reaches greater extents and exhibits more pronounced morphological deviations. The model also produces a clearer asymmetry in the synchrotron brightness of the two lobes than in the baseline case. As anticipated (see \autoref{sec:PrecMod}), overlaying the colour-scaled convolved image from this simulation with the LoTSS DR3 contours of 3C\,306 at 5 and 10 mJy beam$^{-1}$ reveals a systematic misalignment between the northeastern lobe emission and its simulated counterpart.

This interpretation is also supported by the quantitative morphological comparison presented in Appendix~\ref{app:morphological_comparison}. The Intersection over Union (IoU), which measures the common emitting area relative to the total area occupied by either the observed or simulated emission, is ${\rm IoU}=59.5\%$ for the pure precession model, while the Dice coefficient ($D$), which provides a complementary measure of the overall overlap between the two emitting regions, reaches $D=74.6\%$. These values remain close to the corresponding ${\rm IoU}=60.6\%$ and $D=75.5\%$ obtained for the baseline control model. The observational coverage fraction ($f_{\rm obs,covered}$), representing the fraction of the observed radio-emitting area reproduced by the simulation, also remains essentially unchanged, increasing slightly from $73.6\%$ in the baseline model to $74.0\%$ in the pure precession case. At the same time, the simulation excess fraction ($f_{\rm sim,excess}$), which quantifies the fraction of simulated emission extending beyond the observed radio-emitting region, increases from $22.6\%$ to $24.9\%$. This behaviour is consistent with the precessing jet redistributing the emitting material over a broader projected area and producing the larger-scale departures from the mean source axis described above.

These discrepancies suggest that jet precession alone is insufficient to account for the detailed morphology of 3C\,306, motivating the inclusion of additional physical mechanisms, as explored in the following section.

\begin{figure*}
\centering
\includegraphics[width=\textwidth]{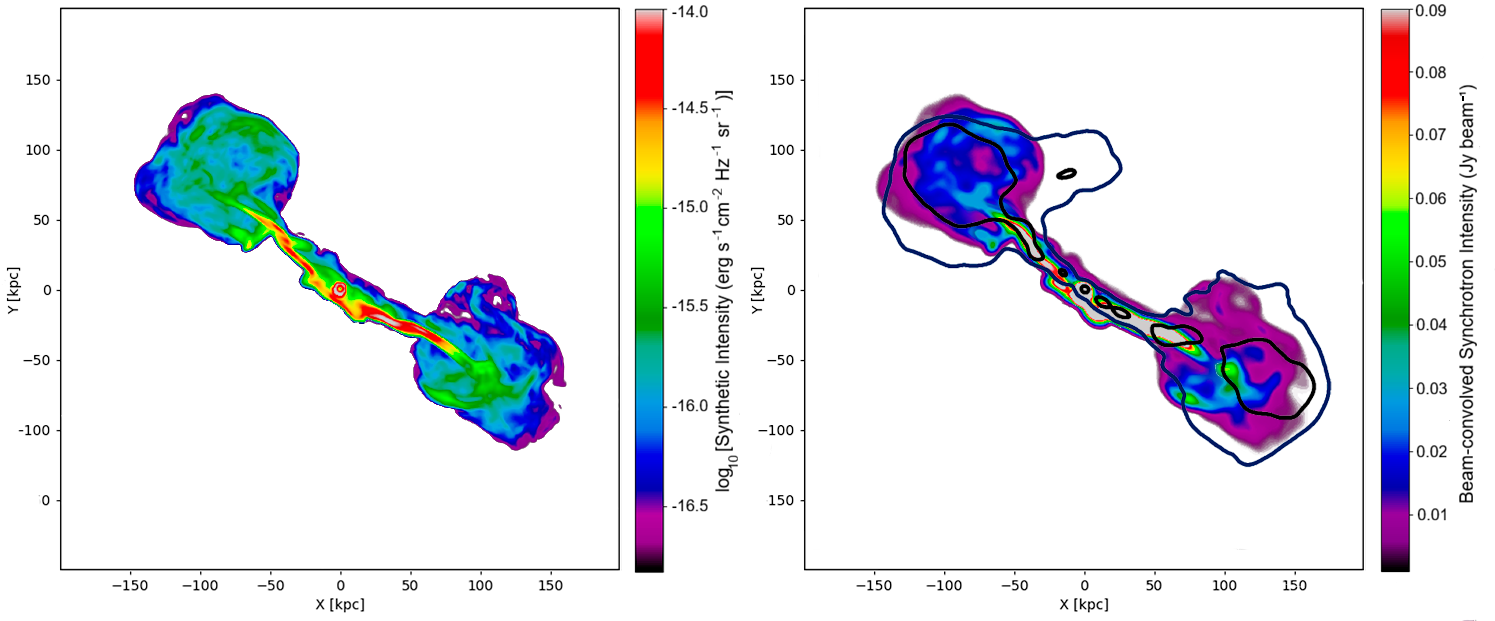}
\caption{Synthetic three-dimensional synchrotron intensity maps for the pure galactic motion case. In the left panel, the projected synchrotron intensity distribution is shown. In the right panel, the corresponding map convolved with a $6.0\arcsec \times 6.0\arcsec$ Gaussian beam and sampled at $0.6702\arcsec$ pixel$^{-1}$, matching the LoTSS DR3 144-MHz observation, is overlaid with observational contours at 0.005 Jy\,beam$^{-1}$ (dark blue) and 0.01 Jy\,beam$^{-1}$ (black).}
\label{fig:mhdresult2}
\end{figure*}

\begin{figure}
\centering
\includegraphics[width=0.47\textwidth]{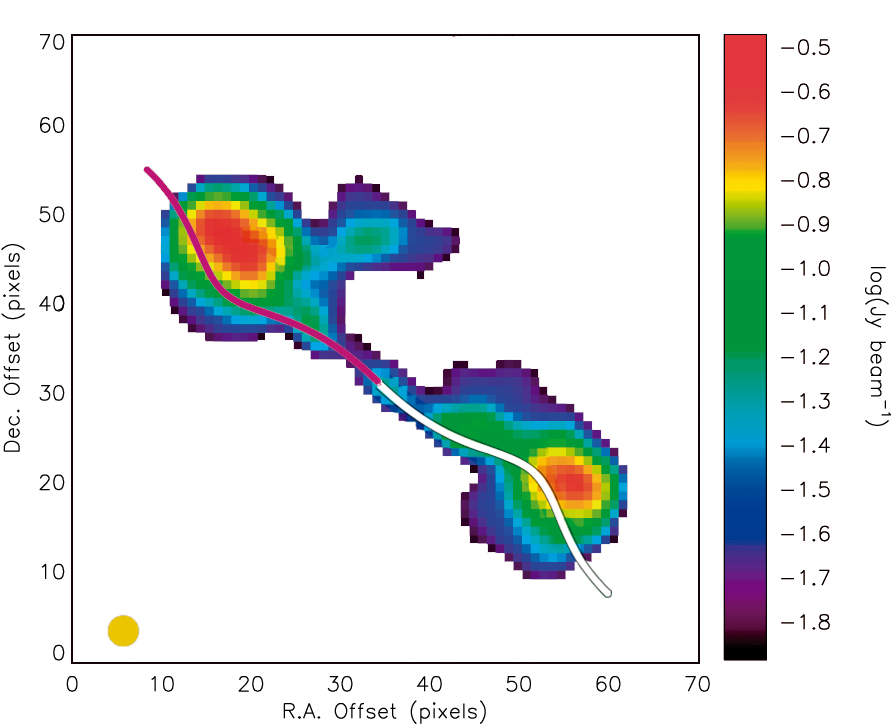}
\caption{Superposition of the modelled precession helices of the jet (white) and counter-jet (purple) over an interferometric image of 3C$\,$306 at 1.4 GHz. The curves represent an 250 Myr precession period, adopting the parameters $\tau_{0,s} = 0.49$, $\phi_0 = 50^{\circ}$, $\varphi_0 = 5^{\circ}$, $\eta_0 = 234^{\circ}$, $\iota = 1$, and $\beta = 0.0034$. The yellow ellipse in the lower-left corner indicates the FWHM of the synthesised CLEAN beam.
}
\label{result_prec_model2}
\end{figure}

\begin{figure*}
\centering
\includegraphics[width=\textwidth]{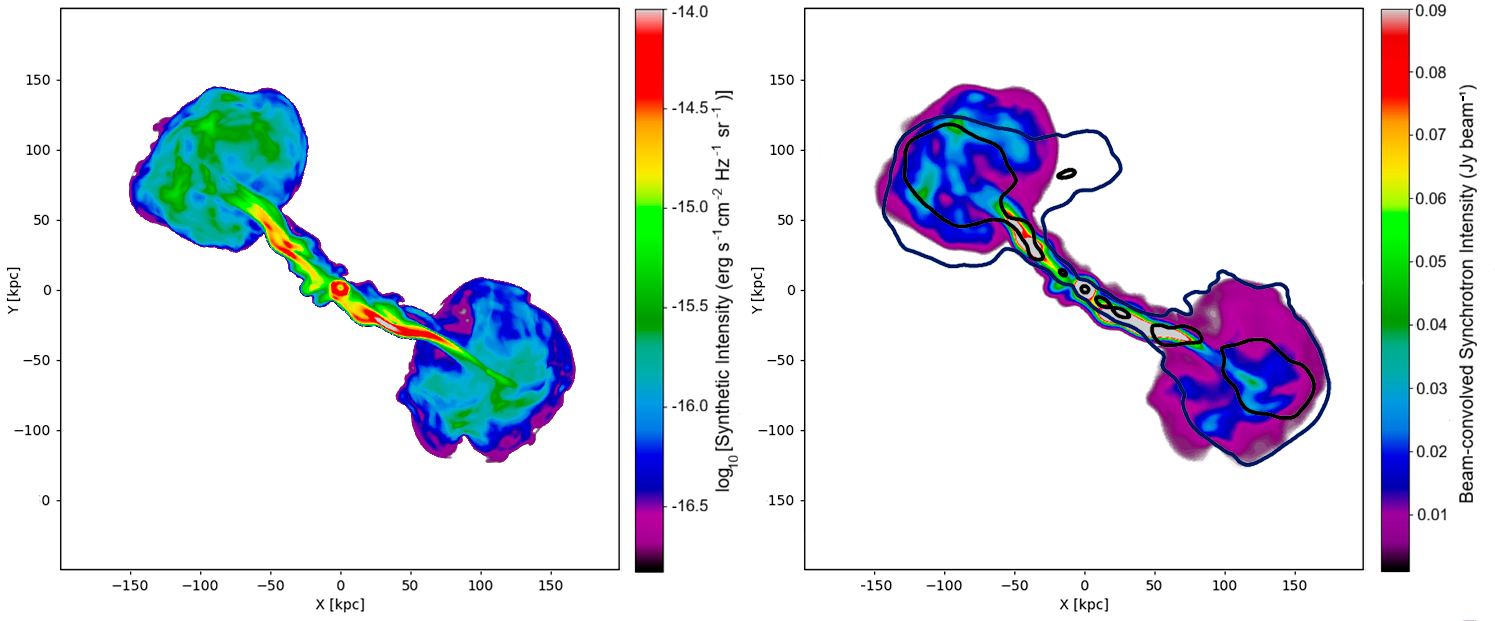}
\caption{
Synthetic three-dimensional synchrotron intensity maps for the composite Model Case. In the left panel, the projected synchrotron intensity distribution is shown. In the right panel, the corresponding map convolved with a $6.0\arcsec \times 6.0\arcsec$ Gaussian beam and sampled at $0.6702\arcsec$ pixel$^{-1}$, matching the LoTSS DR3 144-MHz observation, is overlaid with observational contours at 0.005 Jy\,beam$^{-1}$ (dark blue) and 0.01 Jy\,beam$^{-1}$ (black).
}
\label{fig:mhdresult3}
\end{figure*}

\subsection{Pure galactic motion case}\label{pmc}

\citet{2024MNRAS.533.1341A} showed that the motion of the radio source 3C\,338 relative to its surrounding environment can significantly influence its radio morphology. There are indications that 3C\,338 might be moving with respect to the ICM as well. As discussed in \autoref{subsec:SpectralAge}, the bridge region between the lobes, particularly toward the south-east, exhibits comparatively younger spectral ages over a significant fraction of its extent. This behaviour may reflect local particle re-acceleration or compression, possibly associated with a shock front driven by the source motion. In addition, the discrepancy between the northeastern lobe and the counter-jet precession path seen in \autoref{result_prec_model1} might also be related to this dynamical scenario. The curvature of the jet relative to the longitudinal axis defined by the lobes provides strong evidence for this scenario. It is shown in \autoref{shift} the displacement between the jet inlet region and the longitudinal direction defined by the brightest points of the lobes, identifying a shift in the plane of the sky of about 21.89" or 19.23 kpc at the source's redshift. This offset, combined with a timescale of $\sim 250$ Myr, suggests a projected velocity of $v_{\rm src} \sim 75.2$ km s$^{-1}$ towards the south-east direction. In this case, the central injection position of the jet strictly follows the precomputed orbit ($v_{\rm src} \sim 75.2$ km s$^{-1}$ towards the south-east direction), with no precession applied to the jet direction. 

The synchrotron intensity map obtained from our MHD simulations for the pure galactic motion case is shown in \autoref{fig:mhdresult2}. A visual comparison with the reference image indicates a noticeably improved agreement relative to the pure precession model. In particular, the bridge connecting the two lobes is considerably more collimated and follows the observed emission, suggesting that motion relative to the ICM plays an important role in shaping the source morphology. The simulation also produces a clear morphological asymmetry between the two lobes, while the emission along the bridge is systematically displaced with respect to the intensity peak, indicating that the bulk motion introduces a preferential direction in the plasma distribution.

The quantitative morphological comparison presented in Appendix~\ref{app:morphological_comparison} supports this visual assessment. The pure galactic motion model reaches ${\rm IoU}=63.7\%$ and $D=77.8\%$, improving upon both the baseline control and pure precession ($59.5\%$ and $74.6\%$) models. Most notably, the simulation overlap fraction increases to $f_{\rm sim,overlap}=90.7\%$, while the simulation excess fraction decreases to only $f_{\rm sim,excess}=9.3\%$, indicating that most of the emitting area generated by the moving-source model is contained within the observed radio morphology. However, the observational coverage fraction decreases to $f_{\rm obs,covered}=68.2\%$, showing that the improved confinement of the simulated emission does not imply complete reproduction of the observed source extent.

Despite the overall good agreement, some discrepancies between the simulated and observed emission distributions persist. In the north-western lobe, the simulated emission is slightly displaced northward relative to the observed contours, leaving a substantial portion of the southern part of the observed lobe not reproduced by the simulation. The south-western lobe shows a better overall correspondence in the spatial distribution of the emission; however, the simulated structure does not fully fill the area enclosed by the observed contours, particularly toward the outer regions of the lobe. Although galaxy motion captures the main morphological features of the radio source, additional physical processes, such as jet precession or environmental inhomogeneities, may contribute to a closer match with the observed morphology.

\subsection{Composite Model Case}\label{composite}

In order to provide a more balanced representation of the large-scale radio morphology, we also considered a second jet precession model with $\eta_0 = 234^{\circ}$ and keeping the remaining precession parameters unaltered, as shown in \autoref{result_prec_model2}.
This choice effectively shifts the helices, resulting in an averaged alignment with both lobes, whereas the previous configuration (\autoref{result_prec_model1}) favoured the southwestern lobe. The modification is also motivated by the likely motion of the source relative to the surrounding medium towards the south-east direction.

The implementation of the source motion relative to the ICM in this configuration follows the same prescription described in \autoref{sec:PurePrecMod} and \autoref{pmc}. In practice, this introduces a systematic displacement of the jet inlet region with respect to the longitudinal axis of the lobes, consistent with the evidence discussed previously.

By combining jet precession with the bulk motion of the host galaxy, the composite model provides the closest overall agreement with the observed morphology of 3C\,306 among the scenarios explored in this work, as shown in \autoref{fig:mhdresult3}. A comparison with the LoTSS DR3 reference image (\autoref{fig:ic4516_lotss_ref}) shows that the bridge connecting the two lobes follows the observed emission over most of its extent. The south-western lobe also exhibits an improved spatial correspondence compared with the previous models, which is consistent with the broader redistribution of material introduced by jet precession.

This improvement is also reflected in the quantitative morphological comparison presented in Appendix~\ref{app:morphological_comparison}. The composite model reaches ${\rm IoU}=66.6\%$ and $D=80.0\%$, the highest values among all four scenarios, while its observational coverage fraction increases to $f_{\rm obs,covered}=77.5\%$, also the highest value obtained in our simulations. At the same time, $f_{\rm sim,overlap}=82.6\%$ indicates that most of the simulated emitting area remains confined within the observed radio morphology. Compared with the pure galactic motion model, the inclusion of precession increases the observational coverage from $68.2\%$ to $77.5\%$, although at the expense of a reduction in the simulation overlap fraction from $90.7\%$ to $82.6\%$. This balance results in the highest overall morphological correspondence among the scenarios considered, quantitatively supporting the visual agreement seen in \autoref{fig:mhdresult3}.

Although some discrepancies remain, including regions of the north-western lobe that are not fully filled by the simulated emission within the observational contours and a few detached emission features south of the bridge without clear observational counterparts, the composite model provides the best overall representation of the radio morphology of 3C\,306. The remaining differences may arise from the various simplifications and assumptions adopted in the present modeling framework, and may indicate that additional physical processes or a more complex treatment of the jet–environment interaction could contribute to shaping the observed source.

\section{Discussion}\label{sec:Discussion}

This work combines radio spectral analysis, X-ray constraints on the ICM, and MHD simulations to investigate the origin of the distorted morphology of 3C\,306.

The X-ray observations indicates that RXCJ1454.4+1622 hosts a relatively dense and structured ICM, with a temperature profile suggestive of a cool-core system. Such environments are known to influence the propagation of relativistic jets by modifying pressure gradients and introducing asymmetries in the external medium \citep[e.g.][]{2006PhR...427....1P,2010A&A...513A..37H,2007ARA&A..45..117M}. Although the adopted $\beta$-model provides only a first-order description of the gas distribution, it captures the large-scale conditions relevant for the jet--ICM interaction, which are the dominant factors at the spatial scales probed by the radio emission.

Within this environment, the radio structure of 3C\,306 exhibits clear deviations from axial symmetry, including bent jets and distorted lobes. The spectral index map reinforces this picture by revealing a non-trivial spatial distribution of relativistic electrons, with a curved region of flatter spectra embedded within steeper lobe emission. This behaviour suggests that the jet axis has changed over time, and is consistent with dynamical scenarios involving, for instance, jet precession and interaction with large-scale ICM flows. The jet axis appears to be approximately aligned with the minor axis of IC\,4516, the cD galaxy that hosts 3C\,306. While early works suggested both major-axis alignment and random orientations (e.g., \citealt{1971MNRAS.151..421M,1985ApJ...291...32B}), subsequent analyses have shown that extended radio sources, particularly those with sizes $\gtrsim 200$ kpc (as the case of 3C\,306), tend to exhibit a preferential alignment with the optical minor axis of their host galaxies \citep[e.g.,][]{1989ApJ...336..681B,2009ApJ...695..156S,2015MNRAS.449..955M}. This behaviour is commonly interpreted in terms of anisotropic pressure gradients within the host galaxy, where the minor axis represents the path of least resistance for jet propagation \citep{2009ApJ...695..156S,2015MNRAS.449..955M}. In such a scenario, jets aligned with the minor axis can expand more efficiently and reach larger scales more rapidly, which is supported by studies of giant radio galaxies and their environments \citep{1996MNRAS.279..257S}. However, significant scatter around this trend is observed, and projection effects further complicate the interpretation \citep{1985MNRAS.212..767B,1987MNRAS.229...15S}, indicating that jet orientation is likely influenced by a combination of environmental conditions, accretion history, and possible jet reorientation mechanisms \citep{2012MNRAS.420L...8H,2018MNRAS.474L..81L}. Note that our simulations do not consider the potential contribution of the gas residing within the host galaxy of 3C\,306. This might introduce additional resistance to the jet’s propagation and consequently cause further disturbances to its morphology (e.g., see \citealt{2026arXiv260405471G} for a systematic study of the effects of a triaxial gas distribution on the propagation direction of a kpc-scale jet with power similar to that of 3C\,306).

The spectral ageing analysis provides an additional, independent constraint on this evolution. The relatively uniform distribution of $t_{\rm spec}$ for 3C\,306, with only mild spatial variations, is observed in \autoref{fig:spectralage}. Departures from the simple expectation of a monotonic ageing trend with distance from the core might indicate that the spectral properties of the relativistic electrons are shaped by a combination of radiative losses and additional physical effects. The smooth spectral-age distribution suggests a relatively coherent large-scale evolution of the electron population. However, the limited spatial resolution and frequency coverage imply that the present maps are primarily sensitive to beam-averaged properties, and more complex or time-dependent injection histories cannot be excluded.

An important source of systematic uncertainty in the spectral age estimates arises from the dependence of $t_{\rm spec}$ on both the break frequency and the assumed magnetic field strength (\autoref{eq-tspec}). In particular, $t_{\rm spec} \propto \nu_{\rm br}^{-1/2}$, implying that uncertainties in $\nu_{\rm br}$ directly translate into uncertainties in the inferred ages. Given the limited frequency coverage of the available data, especially at frequencies above $\sim 5$ GHz, the highest break frequencies may not be fully constrained, which can lead to an overestimation of the spectral ages in regions where $\nu_{\rm br}$ lies beyond the sampled range.

In addition, the dependence on the magnetic field strength introduces a non-linear effect through the factor $B^{1/2}/(B^2 + B_{\rm CMB}^2)$. For the equipartition value adopted in this work ($B_{\rm eq} = 2.1\,\mu$G), which is comparable to the inverse-Compton equivalent field $B_{\rm CMB}$ at the source redshift, the resulting ages are close to the maximum values allowed by the model. Deviations from this assumption can significantly affect the inferred timescales.
For instance, magnetic field strengths of 0.1 or 10 $\mu$G can reduce $t_{\rm spec}$ by a factor of $\sim0.3$. Therefore, the absolute values of $t_{\rm spec}$ should be interpreted with caution, and are best regarded as order-of-magnitude estimates that are primarily useful for tracing relative variations across the source. This effect may also contribute to the absence of regions with very low spectral ages in the present analysis. Indeed, spectral ages of about 170 Myr at the core region of 3C\,306 challenges the usual assumption that fresh electrons are continuously injected at the core region during the whole evolution of 3C\,306. Instead, this galaxy might have undergone multiple episodic outbursts, as commonly detected in the Double-Double Radio Galaxies (e.g., \citealt{2000MNRAS.315..371S, 2009BASI...37...63S, 2012BASI...40..121N, 2013MNRAS.436.1595K, 2006MNRAS.366.1391S, 2017MNRAS.467L..56N, 2026Galax..14....2G}), being the core region still embedded in the remnant plasma of a previous epoch. Its kpc-scale nucleus exhibits current AGN activity, corroborated by its continuum emission detected from radio to X-ray observations. These features support a possible intermittence in the jet activity of this radio galaxy during the last hundreds of Myr.

An additional morphological feature is that the jet/bridge structure appears to enter the lobes approximately through their central regions, rather than connecting preferentially to their outermost edges. This is particularly evident in the \autoref{fig:ic4516_lotss_ref} and is also consistent with the spectral-age map (\autoref{fig:spectralage}), where the bridge is associated with relatively younger plasma that extends continuously into the inner parts of both lobes. Such a configuration suggests that the present large-scale radio morphology is not simply the result of two disconnected lobe structures linked by a later, unrelated jet episode, but instead reflects a dynamical connection between the bridge and the lobe interiors over a substantial fraction of the source lifetime.

This behaviour provides an additional constraint on the simulated scenarios when the location at which the jet/bridge enters the lobes is considered together with the position of the brightest lobe regions. A comparison between the synthetic synchrotron maps and the \autoref{fig:ic4516_lotss_ref} shows that the pure galactic motion and composite models provide the closest qualitative correspondence in this respect. In both cases, the synchrotron-emitting channel enters the lobes through their inner/central regions while remaining approximately connected to the brightest emission associated with the lobe termination regions. In the pure precession model, by contrast, the structures produced near the bridge--lobe transition are systematically displaced from the brightest observed emission, an effect that is particularly evident on the south-western side. Thus, although precession can redistribute the emitting plasma and reproduce some of the large-scale morphological characteristics of the source, source motion appears to be required to simultaneously reproduce the location of the jet entry into the lobes and its spatial relation to the brightest lobe regions. The composite model retains this correspondence while allowing the additional redistribution of emitting material introduced by precession.

While the radio images and spectral-age map of 3C\,306 suggest a complex dynamical history, they do not uniquely constrain the physical processes responsible for the observed morphology. The comparison with the MHD simulations indicates that the pure precession model reproduces some of the large-scale characteristics of the source, but tends to overestimate the transverse extent of the bridge and does not satisfactorily reproduce its observed alignment. It also fails to recover the characteristic curvature of the jet/bridge structure seen both in the \autoref{fig:ic4516_lotss_ref} and in \autoref{shift}. This curvature is reproduced only when source motion is included, being present in both the pure galactic-motion and composite models. 

In the pure source-motion case, the bulk displacement of the host galaxy relative to the ICM naturally produces a bent and well-aligned bridge while substantially reducing the excess of simulated emission, providing a improved representation of the observed radio structure. The composite model preserves this characteristic curvature while introducing the additional redistribution of emitting material associated with precession, and yields the highest overall morphological agreement among the scenarios explored here. 

The fact that the observed curvature is recovered in both models that include source motion, but not in the pure precession case, provides further support for the motion of the host galaxy through the ICM as the dominant mechanism shaping the large-scale morphology of 3C\,306. Jet precession therefore appears to play a secondary role, further improving the overall correspondence with the observations when combined with source motion.

In this context, it is important to emphasize that the dynamical scenario explored in this work relies on a simplified geometrical assumption for the motion of the host galaxy. In particular, the adopted orbital path is approximated as a linear trajectory confined to the plane of the sky, which should be regarded as an idealised representation of a potentially more complex three-dimensional orbit. Under this assumption, the inferred source velocity of $\sim 75.2$ km\,s$^{-1}$ is derived by adopting a characteristic timescale of $\sim 250$ Myr, consistent with the spectral age estimates in the radio lobes and the assumed orbital geometry. As such, this velocity is not a direct observable, but rather depends on the combination of these modelling choices. If the true orbit is inclined with respect to the plane of the sky, or if the relevant dynamical timescale is shorter than 250 Myr, the implied orbital velocities would be correspondingly higher. In such cases, the required velocities could approach or exceed the typical peculiar velocities observed for central cluster galaxies, which are generally of the order of a few hundred km\,s$^{-1}$ \citep[e.g.,][]{2015ApJ...807...56B,2009AJ....137.4795C}, corresponding to $\sim 30\%$ of the cluster velocity dispersion. Although larger values have been reported in individual systems (e.g., up to $\sim 700$ km\,s$^{-1}$; \citealt{2006MNRAS.368..651P}), these are typically associated with dynamically disturbed environments and are not representative of relaxed clusters. 
Therefore, the inferred velocity should be interpreted as a lower-limit estimate within the framework of the adopted assumptions.

An alternative interpretation of the apparent displacement between the radio structure and the surrounding medium is that the observed morphology may not be solely driven by the motion of the host galaxy within the cluster potential. Instead, it is possible that the central supermassive black hole itself is offset from the dynamical centre of the galaxy. In this scenario, the jet orientation and large-scale morphology would reflect the motion of a wandering black hole within the galactic potential, rather than the bulk motion of the entire galaxy. Such offsets may arise from past merger events or dynamical interactions, leading to oscillations of the black hole around the potential minimum (e.g., \citealt{2023MNRAS.522..948C,2021ApJ...916L..18R}). Although the present data do not allow us to distinguish unambiguously between these scenarios, the possibility of a wandering black hole provides an alternative framework to interpret the observed asymmetries without requiring large-scale motion of the host galaxy. Future high-resolution observations of the central region would be required to test this hypothesis.

\section{Conclusions} \label{sec:conclusions}

We investigated the origin of the distorted radio morphology of 3C\,306 by combining radio spectral analysis, X-ray constraints on the ICM, and three-dimensional MHD simulations.

The main results of this work can be summarised as follows:

\begin{enumerate}
\item {\bf X-ray properties of the galaxy cluster RXCJ1454.4+1622.} The temperature profile derived from \textit{XMM--Newton} data indicates a low-mass cool-core system, while the density distribution inferred from a $\beta$-model provides a first-order description of the ICM. These profiles establish the environmental conditions used as input for the jet simulations.

\item {\bf Radio spectral analysis of 3C\,306.} Break-frequency mapping reveals a relatively uniform spectral curvature across the source, with modest spatial variations. Independent JP modelling of the lobe-integrated spectra yields break frequencies of a few hundred MHz, broadly consistent with the spatially resolved analysis.

\item {\bf Radiative ages and magnetic field.} Assuming equipartition between the magnetic and the total particle energies, we derive a characteristic magnetic field of about $2.1\,\mu$G and spectral ages of order $220$ Myr. These values should be interpreted as effective, beam-averaged quantities, subject to systematic uncertainties associated with the magnetic field strength, spectral modelling assumptions, and limited frequency coverage.

\item {\bf  Jet--ICM interaction from MHD simulations.} Pure precession modifies the propagation and brightness distribution of the jet/counter-jet system but does not reproduce the overall observed morphology. In contrast, source motion accounts for the main large-scale morphological features, indicating that the motion of the host galaxy through the ICM is the dominant mechanism shaping the radio structure. The composite model provides the highest overall morphological agreement among the scenarios explored, with jet precession playing a secondary role in improving the correspondence with the observations.
\end{enumerate}

An important aspect of this work is the exploration of the {\it combined} effect of jet precession and bulk motion of the source within the ICM. While these mechanisms have been extensively investigated individually in previous numerical studies (e.g., \citealt{1988ApJ...327..627S, 1992ApJ...393..631B, 2010ApJ...713L..74F, 2016MNRAS.458..802N, 2017ApJ...839...14G, 2020MNRAS.499.5765H, 2022ApJ...933...71F, 2022MNRAS.514.5625G, 2024MNRAS.533.1341A}), simulations explicitly coupling both effects within a single dynamical framework remain relatively rare in phenomenological studies. In this context, the present work provides a novel, phenomenological approach to assess how the interplay between intrinsic jet dynamics and environmental interaction shapes the observed morphology of radio galaxies.

Although the results are internally consistent within the adopted assumptions, several limitations must be considered. Departures from equipartition energy condition, spatial variations in the physical parameters of the ICM, or the presence of re-acceleration processes may affect the inferred spectral ages. Likewise, the simulated morphologies do not reproduce all observed features in detail, indicating that additional physical ingredients — such as ICM inhomogeneities or time-dependent jet activity — may play a role.

In summary, the observational analysis and simulations suggest that the large-scale morphology of 3C\,306 is shaped by a combination of intrinsic jet dynamics and environmental interaction. Although the relative importance of jet precession and motion relative to the ICM cannot be uniquely constrained within the present modelling framework, the results indicate that both effects likely contribute to the observed structure. Future observations with broader radio frequency coverage, deeper X-ray data, and more realistic modelling of the ICM will be important to further refine these constraints. Nevertheless, this work demonstrates that combining multi-wavelength observations with controlled MHD simulations provides a physically motivated framework for investigating the evolution of radio galaxies in cluster environments.

\section*{Acknowledgements}

The authors acknowledge the National Laboratory for Scientific Computing (LNCC/MCTI, Brazil) for providing HPC resources from the SDumont supercomputer, which contributed to the research results reported in this paper. URL: \url{http://sdumont.lncc.br}. ASRA and TFL acknowledge financial support from the Conselho Nacional de Desenvolvimento Científico e Tecnológico (CNPq) through grants 684846/2022-00 and Productivity Fellowship 303672/2022-6, respectively. L.L. acknowledges support from DGAPA/UNAM PAPIIT grants IN108324 and IN112820, CONAHCyT grant CF-263356, and SECIHTI grant CBF-2025-I-109. The authors thank the anonymous referee for the constructive and insightful comments, which significantly improved the clarity and overall quality of the manuscript.

\section*{Data Availability}

The data generated in this research will be shared on reasonable request to the corresponding author.



\bibliographystyle{mnras}
\bibliography{references} 




\appendix

\section{Baseline control simulation}
\label{app:baseline_model}

\begin{figure*}
\centering
\includegraphics[width=1\textwidth]{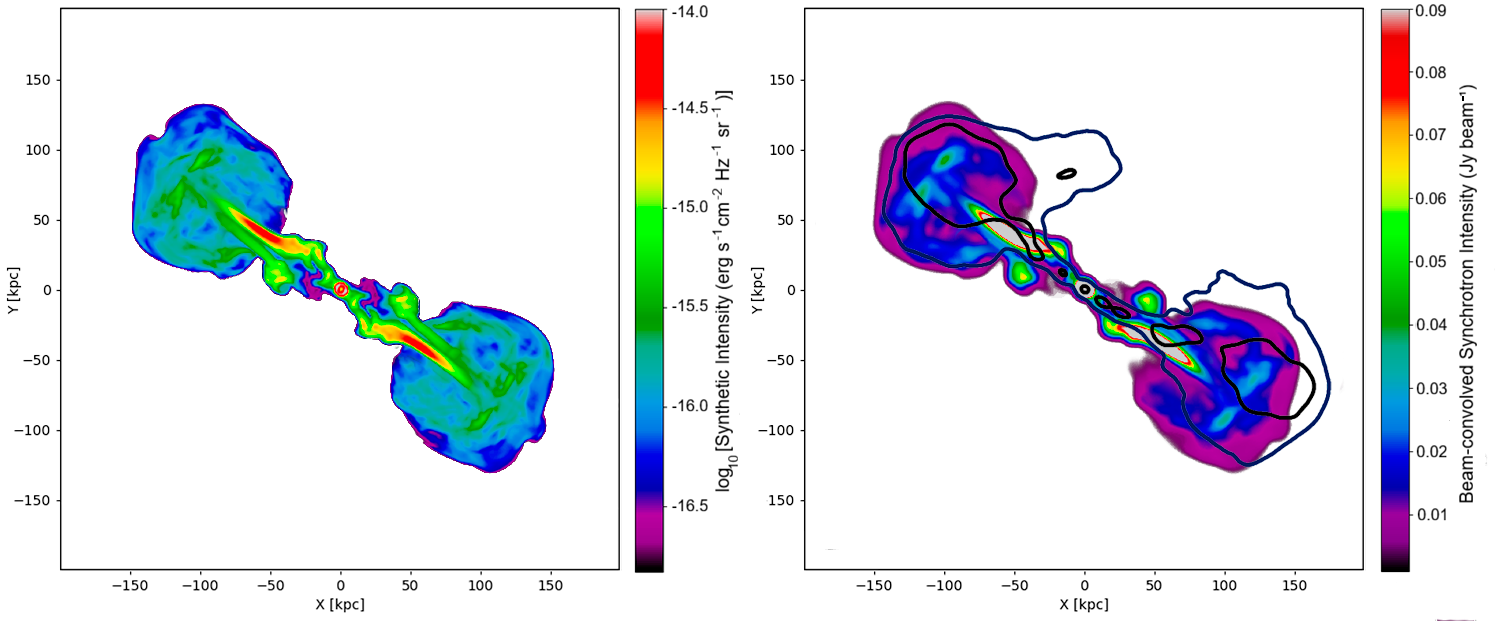}
\caption{Synthetic three-dimensional synchrotron intensity maps for the baseline control simulation. In the left panel, the projected synchrotron intensity distribution is shown. In the right panel, the corresponding map convolved with a $6.0\arcsec \times 6.0\arcsec$ Gaussian beam and sampled at $0.6702\arcsec$ pixel$^{-1}$, matching the LoTSS DR3 144-MHz observation, is overlaid with observational contours at 0.005 Jy\,beam$^{-1}$ (dark blue) and 0.01 Jy\,beam$^{-1}$ (black).}
\label{fig:baseline_control}
\end{figure*}

We run an additional control simulation with a non-precessing and static jet inlet region at the centre of the computational domain, where is imposed the same physical conditions adopted in the simulations listed in \autoref{tab:jetparams}.  This run preserves the oblique jet orientation relative to the computational grid, thereby isolating the effects of the tilted jet/counter-jet injection. \autoref{fig:baseline_control} shows the synchrotron emissivity integrated along the line of sight (left panel) and its respective convolved synchrotron map of this baseline control simulation, together with the observational radio contours of 3C\,306 at 144 MHz (right panel).

In the absence of both jet precession and bulk motion of the jet inlet region, the simulated radio structure remains predominantly distributed along the fixed jet injection direction. However, the jet and counter-jet do not evolve as a perfectly straight or axisymmetric flow. Small oscillations are visible along the bridge connecting the core and the lobes. We interpret these features as a natural consequence of the interaction between the light jet and the surrounding ICM. As the flow propagates, it continuously excavates and subsequently travels through its own cocoon, so that the interaction between the jet, the backflow, and the surrounding medium can generate small deflections of the beam. Such perturbations may also be amplified by the development of hydrodynamic and magnetohydrodynamic instabilities, including Kelvin--Helmholtz and current-driven kink modes (e.g. \citealt{2006A&A...456..493P,1998ApJ...493..291B,2000A&A...355..818A,2017MNRAS.469.4957B}). Indeed, the injection of a purely toroidal magnetic field makes the jet prone to develop wiggling patterns due to kink instabilities (e.g., \citealt{2016MNRAS.456.1739B, 2012ApJ...757...16M, 2017A&A...606A..57R}). Note also that the combination of a jet injection angled to the grid $z-$axis the method to deal numerically with the divergence of the magnetic field might also introduce additional instabilities and filamentary structures at the jet inlet region, even though they are possibly small in amplitude given the low magnetic-dominated regime of the MHD simulations in this work (thermal pressure of the jet is roughly four times larger than its magnetic pressure at the jet inlet region).

The comparison between the convolved simulated image and the 144-MHz contours displayed in \autoref{fig:baseline_control} indicates that the oblique injection geometry alone is insufficient to reproduce the smooth bent morphology of 3C\,306. In the absence of both precession and bulk source motion, the simulated radio emission remains predominantly distributed along the fixed jet injection direction, but clear morphological discrepancies persist. While the northeastern lobe shows a reasonable overall correspondence with the observed radio structure, larger deviations are found in the bridge regions approaching the lobes and, particularly, in the southwestern lobe. The characteristic curvature of the observed bridge--lobe transitions and the morphology and extent of the southwestern emission are not adequately reproduced by the baseline model.

The additional broadening of the jet channel close to core ($r<30$ kpc) due to precession and bulk motion is evident. While our control image exhibits a transverse width of about 14 kpc, the pure precession, pure galactic motion, and the composite model cases present respectively a jet broadening of about 22, 27 and 29 kpc.

\begin{figure}
    \centering
    \includegraphics[width=\columnwidth]{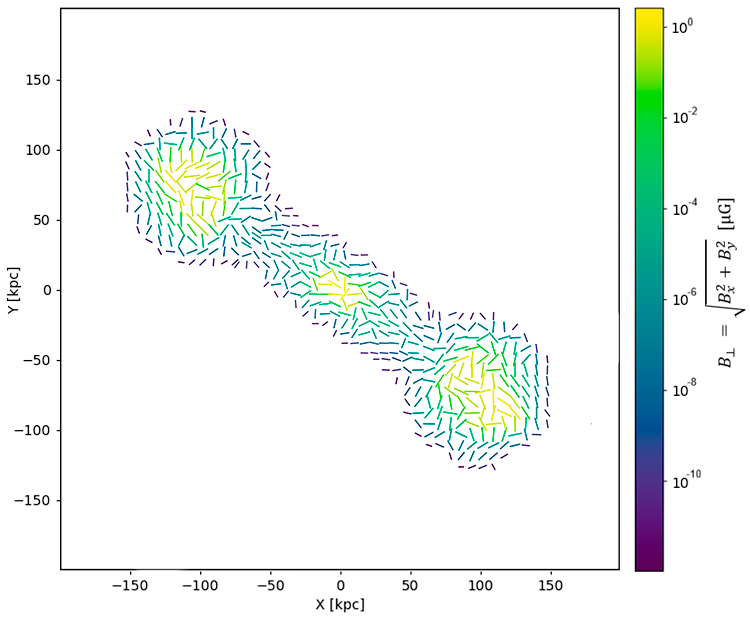}
    \caption{Projected magnetic-field structure of the baseline control simulation in the $xy$ plane. Line segments indicate the local orientation of the projected magnetic field, $(B_x,B_y)$, without distinguishing its polarity, while the colour scale with the projected field strength, $B_{\perp}=(B_x^2+B_y^2)^{1/2}$.}
    \label{fig:baseline_magnetic_field}
\end{figure}

The projected magnetic-field distribution provides an additional diagnostic of the internal structure of the baseline model. As shown in \autoref{fig:baseline_magnetic_field}, the projected field strength exhibits an approximately symmetric distribution between the two sides of the source. The strongest fields are concentrated near the central region of the bridge and within the inner regions of both lobes, whereas the field strength generally decreases towards the outer boundaries of the radio structure. The projected field orientations show a more complex pattern within the lobes, reflecting the redistribution of the magnetic field as the jet material expands and circulates within these regions. Overall, the approximate bilateral symmetry of both the field-strength distribution and the large-scale magnetic structure is consistent with the symmetric jet/counter-jet injection adopted in the baseline model.

\section{Quantitative comparison between synthetic and observed radio emission}
\label{app:morphological_comparison}

To quantify the morphological agreement between the observed radio emission and the synthetic synchrotron maps of 3C\,306 independently of their absolute surface-brightness distributions, we performed a binary-mask comparison based on the projected spatial extent of the emission. Intensity information is used only to determine whether a given pixel belongs to the radio-emitting region; after the masks are constructed, the comparison depends solely on the spatial distribution and projected area of the observed and simulated emission.

Prior to the comparison, the observational and synthetic maps were placed on a common physical scale and resampled onto the same spatial grid, such that each pixel corresponds to the same projected physical area in both images. Regions associated with unrelated field sources were excluded from the masking procedure and the morphological comparison. The observed radio-emitting region was defined using a threshold of $4\sigma_{\rm rms}$, where $\sigma_{\rm rms}$ is the rms noise of the \autoref{fig:ic4516_lotss_ref}. The observational binary mask, $O_i$, is therefore defined as

\begin{equation}
O_i =
\begin{cases}
1, & S_{{\rm obs},i} > 4\sigma_{\rm rms},\\
0, & S_{{\rm obs},i} \leq 4\sigma_{\rm rms},
\end{cases}
\label{eq:obs_binary_mask}
\end{equation}
\\where $S_{{\rm obs},i}$ is the observed surface brightness at pixel $i$. For the synthetic map, the emitting region was defined relative to the peak synthetic surface brightness,

\begin{equation}
M_i =
\begin{cases}
1, & S_{{\rm sim},i} > f_{\rm th}S_{\rm sim,max},\\
0, & S_{{\rm sim},i} \leq f_{\rm th}S_{\rm sim,max},
\end{cases}
\label{eq:sim_binary_mask}
\end{equation}
\\where $S_{{\rm sim},i}$ is the synthetic surface brightness at pixel $i$, $S_{\rm sim,max}$ is its maximum value, and $f_{\rm th}=0.02$ is the adopted fractional threshold.

The two binary masks divide the analysed region into three mutually exclusive classes: pixels containing emission in both the observation and simulation, pixels containing only observed emission, and pixels containing only simulated emission. The total observed and simulated areas are defined as $A_{\rm obs} = \sum_i O_i$ and $A_{\rm sim} = \sum_i M_i,$ while the common area is $A_{\rm int} = \sum_i O_i M_i$. The regions present exclusively in the observation and simulation are then $A_{\rm obs,only}=A_{\rm obs}-A_{\rm int}$ and $A_{\rm sim,only}=A_{\rm sim}-A_{\rm int},$ respectively.

\subsection{Morphological overlap metrics}
\label{app:morphological_metrics}

The morphological correspondence was quantified using complementary area-based metrics that separately evaluate how much of the simulated emission is consistent with the observation and how much of the observed structure is reproduced by the simulation.

From the simulation perspective, the \textit{simulation overlap fraction} is defined as

\begin{equation}
f_{\rm sim,overlap}
=
100\,
\frac{A_{\rm int}}{A_{\rm sim}},
\label{eq:sim_overlap}
\end{equation}
\\and gives the percentage of the simulated emitting area that lies within the observed radio-emitting region. Its complementary quantity, the \textit{simulation excess fraction},

\begin{equation}
f_{\rm sim,excess}
=
100\,
\frac{A_{\rm sim,only}}{A_{\rm sim}},
\label{eq:sim_excess}
\end{equation}
\\measures the fraction of simulated emission produced outside the observed radio morphology of 3C\,306.

From the observational perspective, the \textit{observational coverage fraction} is

\begin{equation}
f_{\rm obs,covered}
=
100\,
\frac{A_{\rm int}}{A_{\rm obs}},
\label{eq:obs_covered}
\end{equation}
\\which gives the percentage of the observed radio-emitting area reproduced by the simulation. 

Similarly, the complementary \textit{missing-observation fraction},

\begin{equation}
f_{\rm obs,missing}
=
100\,
\frac{A_{\rm obs,only}}{A_{\rm obs}},
\label{eq:obs_missing}
\end{equation}
\\quantifies the fraction of the observed morphology not recovered by the model.

A global measure of morphological agreement was obtained using the Intersection over Union (IoU) metric. Defining the union of the two emitting regions as $A_{\rm union} = A_{\rm obs}+A_{\rm sim}-A_{\rm int},$ the IoU metric can be defined as

\begin{equation}
{\rm IoU}
=
100\,
\frac{A_{\rm int}}{A_{\rm union}}.
\label{eq:iou}
\end{equation}

The IoU simultaneously accounts for both observed emission not reproduced by the simulation and simulated emission extending beyond the observed source. It reaches $100\%$ only when the two binary masks are identical.

As an additional global similarity measure, we calculate the Dice coefficient,

\begin{equation}
D
=
100\,
\frac{2A_{\rm int}}
     {A_{\rm obs}+A_{\rm sim}}.
\label{eq:dice}
\end{equation}

Like the IoU, the Dice coefficient ranges from $0$ to $100\%$, with $100\%$ corresponding to identical masks. For partially overlapping structures, the Dice coefficient generally yields larger values than the IoU because it assigns greater relative weight to the common area.

This analysis therefore quantifies the agreement in projected radio morphology rather than in absolute synchrotron surface brightness. The resulting metrics are sensitive to the position, projected extent, and spatial overlap of the radio-emitting structures, while remaining insensitive to differences in the detailed surface-brightness distribution within those regions. In this way, the geometrical agreement between the simulated and observed radio structures can be evaluated independently of their absolute synchrotron normalisation.

\subsection{Quantitative morphological comparison}
\label{sec:morphological_comparison}

\begin{figure*}
    \centering
    \includegraphics[width=1.0\textwidth]{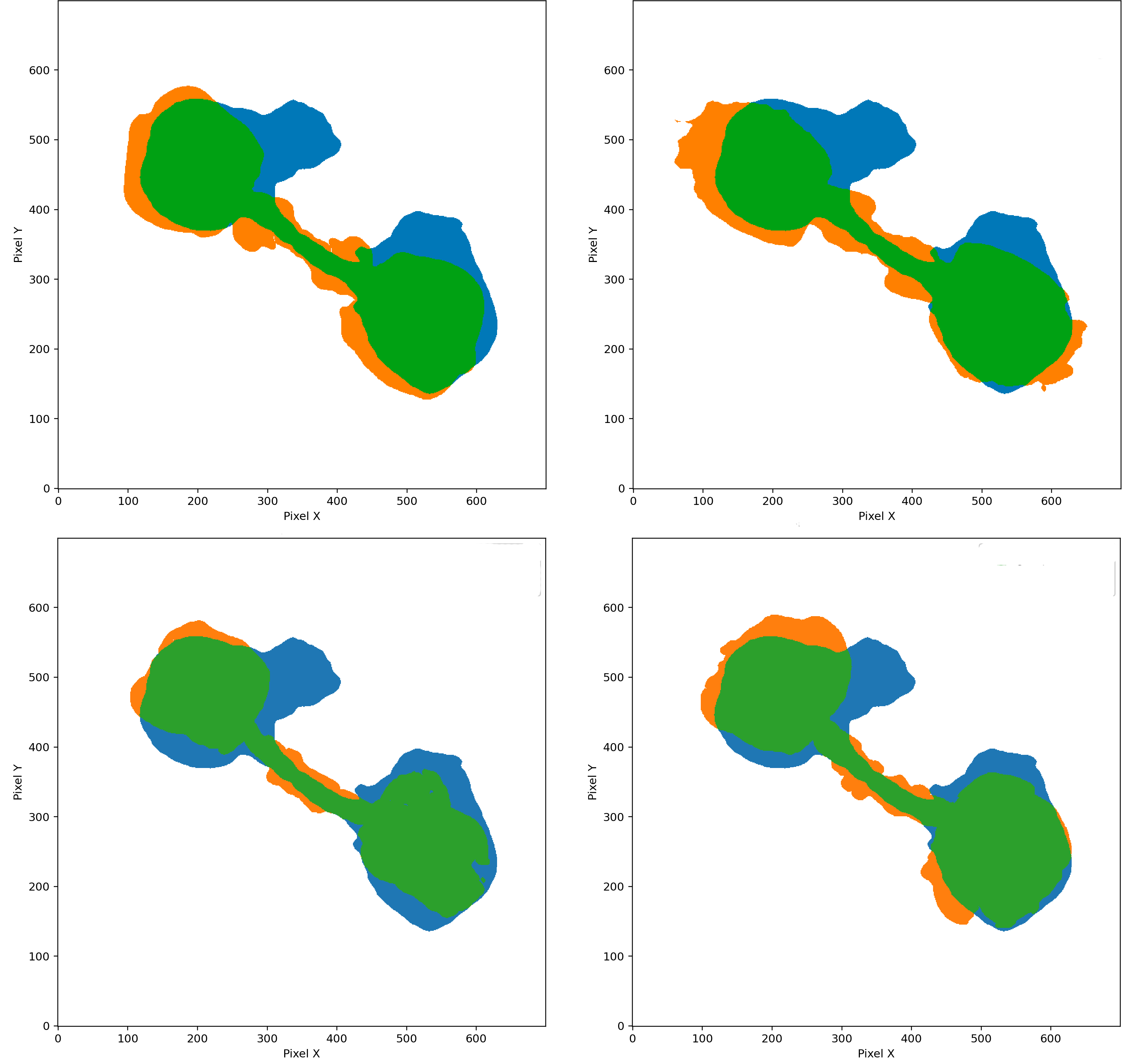}
    \caption{Morphological comparison between the observed LOFAR radio emission and the synthetic emission obtained for the four simulated scenarios. The panels show the baseline control model (top left), pure precession model (top right), pure galactic motion model (bottom left), and composite model (bottom right). Blue regions represent emission detected only in the observation, green regions indicate the spatial overlap between the observed and simulated emission, and orange regions represent emission present only in the simulation.}
    \label{fig:morphological_overlap}
\end{figure*}

The spatial correspondence between the observed and simulated emitting regions for the baseline control, pure precession, pure galactic motion, and composite models are shown in \autoref{fig:morphological_overlap}. The comparison distinguishes regions reproduced by both maps from those present exclusively in either the observation or the simulation.

The resulting morphological metrics are summarised in \autoref{tab:morphological_metrics}. Among the four scenarios, the composite model provides the highest global morphological agreement with the observed radio source, reaching ${\rm IoU}=66.6\%$ and $D=80.0\%$. It also yields the largest observational coverage fraction, $f_{\rm obs,covered}=77.5\%$, and consequently the smallest fraction of observed emission not recovered by the simulation, $f_{\rm obs,missing}=22.5\%$. 

\begin{table*}
\centering
\caption{Morphological comparison between the observed LOFAR radio emission and the synthetic radio emission for the four simulated scenarios. All values are expressed as percentages.}
\label{tab:morphological_metrics}
\begin{tabular}{lcccccc}
\hline
Model &
$f_{\rm sim,overlap}$ &
$f_{\rm sim,excess}$ &
$f_{\rm obs,covered}$ &
$f_{\rm obs,missing}$ &
IoU &
Dice \\
\hline
Baseline control     & 77.4 & 22.6 & 73.6 & 26.4 & 60.6 & 75.5 \\
Pure precession      & 75.1 & 24.9 & 74.0 & 26.0 & 59.5 & 74.6 \\
Pure galactic motion & 90.7 & 9.3 & 68.2 & 31.8 & 63.7 & 77.8 \\
Composite            & 82.6 & 17.4 & 77.5 & 22.5 & 66.6 & 80.0 \\
\hline
\end{tabular}
\end{table*}

The pure galactic motion model provides the second-highest global agreement, with ${\rm IoU}=63.7\%$ and $D=77.8\%$. Interestingly, this model reaches the highest simulation overlap fraction among all scenarios, $f_{\rm sim,overlap}=90.7\%$, together with the lowest simulation excess fraction, $f_{\rm sim,excess}=9.3\%$. Thus, most of the radio-emitting area generated by this model lies within the observed morphology. However, its observational coverage remains $68.2\%$, indicating that nearly one third of the observed emitting area is not reproduced by the simulation.

The baseline control and pure precession models show lower overall morphological correspondence. The baseline model reaches ${\rm IoU}=60.6\%$ and $D=75.5\%$, whereas the pure precession model yields ${\rm IoU}=59.5\%$ and $D=74.6\%$. Their observational coverage fractions are also similar, $73.6\%$ and $74.0\%$, respectively. The pure precession model additionally presents the largest simulation excess fraction ($24.9\%$), indicating a comparatively larger fraction of synthetic emission extending into regions not occupied by the observed radio source.


\bsp	
\label{lastpage}
\end{document}